\documentclass[acmtocl,acmnow]{acmtrans2m}
\acmVolume{4} %the volume
\acmNumber{1} %the number
\acmYear{08}  %the last two digits of the year,
\acmMonth{March}  %the month
\pdfoutput=1
\usepackage{float}
\usepackage{ifthen}
\newboolean{acmformat}
\setboolean{acmformat}{true}   %formatted per acm
\newboolean{evilunits}
\setboolean{evilunits}{true}   %formatted with bad units
\newcommand{\KiB}{\ifthenelse{\boolean{evilunits}}{kB}{KiB}}
\newcommand{\MiB}{\ifthenelse{\boolean{evilunits}}{MB}{MiB}}
\newcommand{\GiB}{\ifthenelse{\boolean{evilunits}}{GB}{GiB}}
\newcommand{\TiB}{\ifthenelse{\boolean{evilunits}}{TB}{TiB}}
\newcommand{\mebibyte}{\ifthenelse{\boolean{evilunits}}{megabyte}{mebibyte}}
\newcommand{\tebibyte}{\ifthenelse{\boolean{evilunits}}{terabyte}{tebibyte}}

\newtheorem{theorem}{Theorem}[section]

\newtheorem{proposition}[theorem]{Proposition}
\newtheorem{lemma}[theorem]{Lemma}
\newdef{definition}[theorem]{Definition}
\newdef{remark}[theorem]{Remark}

\usepackage[T1]{fontenc}
\usepackage[latin1]{inputenc}

\usepackage{amsmath}
\usepackage{amssymb}
\usepackage{ifpdf}
\ifpdf
    \providecommand{\myfig}[1]{#1.pdf}
   \usepackage[pdftex]{graphicx}

\else
   
\providecommand{\myfig}[1]{#1.eps}
   \usepackage{graphicx}

\fi
\usepackage{verbatim}
\usepackage{pslatex}
\usepackage{dsfont}
\usepackage{algorithm}
\usepackage{algorithmic} % can use noend to play with spacing
\usepackage{colortbl}
\definecolor{webdarkgray}{rgb}{0.5,0.5,0.5}
\definecolor{webgray}{rgb}{0.65,0.65,0.65}
\definecolor{weblightgray}{rgb}{0.85,0.85,0.85}

 \newcommand{\owensilent}[1]{}
 \newcommand{\danielsilent}[1]{}

\newcommand{\structure}[1] {} %{ \{\textbf{Document Structure comment: #1}\}}

\newcommand{\singlespace}{\renewcommand{\baselinestretch}{1.0}\small \normalsize}
\title{Hierarchical Bin Buffering: Online Local Moments for Dynamic External Memory Arrays}
\author{DANIEL LEMIRE\\Universit\'e du Qu\'ebec \`a Montr\'eal \and
OWEN KASER\\University of New Brunswick}
\begin{abstract}
  For a massive I/O array of size $n$, we want to
  compute the first $N$ local moments, for some constant $N$.
  Our simpler algorithms partition the array into consecutive ranges
  called bins, and apply not only to local-moment queries,
  but also to algebraic queries. % in general.
    With $N$ buffers of
  size $\sqrt{n}$, time complexity drops to $O(\sqrt n)$.  A more
  sophisticated approach uses hierarchical buffering and has a
  logarithmic time complexity ($O(b \log_b n)$), when using $N$
  hierarchical buffers of size $n/b$. Using Overlapped Bin Buffering, we
  show that only one buffer is needed, as with wavelet-based
  algorithms, but using much less storage.
\end{abstract}



\begin{document}
%\begin{bottomstuff}
\bottomstuff
The first author was supported by NSERC grant~261437 and the second
author was supported by NSERC grant~155967.\\
Authors' addresses: Daniel Lemire, Universit\'e du Qu\'ebec \`a Montr\'eal, 100 Sherbrooke West,
  Montr\'eal, QC H2X~3P2  Canada and
 Owen Kaser,  University of New Brunswick, 100 Tucker Park Road,  Saint John, NB E2L~4L1 Canada
\end{figure}

%\end{bottomstuff}
\maketitle

%\date{\today{}}

%\author{Daniel Lemire         \and
%        Owen Kaser %etc.
%}

%\institute{Daniel Lemire \at
%Universit\'e du Qu\'ebec \`a Montr\'eal (UQAM)\\
%100 Sherbrooke West\\
%Montreal, QC\\
%H2X 3P2 Canada\\
%Tel.: (514) 987-3000 ext. 2835\\
%Fax: (514) 843-2160\\
%              \email{lemire@acm}
%           Owen Kaser \at
%UNB Saint John \\
%P.O. Box 5050 \\
%Saint John, NB \\
%E2L 4L5 Canada\\
%              \email{owen@computer.org}           %  \\
%}

%\newpage

%\date{Received: date / Revised: date}
%\maketitle

%\category{H.3.5}{Information Storage}{Online Information Services}
%\category{G.1.1}{Numerical Analysis}{Interpolation}

%\terms{Algorithms, Theory, Experimentation}
%\keywords{Very Large Arrays, Hierarchical Buffers, Statistical Queries, Polynomial Fitting}

%\begin{keyword}
%Very Large Arrays \sep Hierarchical Buffers \sep Statistical Queries \sep Polynomial Fitting
%\end{keyword}
%\end{frontmatter}
%\newpage
%%%%%%%%%%%%%%%%%%%%%%%%%%%%%%%%%%%%%%%%%%%%%%%%%%%%%%%%%%%%%%%%%%%%%%%%%%%%%%%%%%%%%%%%%%%%%%%%%%
%%%%%%%%%%%%%%%%%%%%%%%%%%%%%%%%%%%%%%%%%%%%%%%%%%%%%%%%%%%%%%%%%%%%%%%%%%%%%%%%%%%%%%%%%%%%%%%%%%

%\begin{bottomstuff}
%Author's address: Universit�du Qu�ec,
% 4750, avenue Henri-Julien,
% Montr�l, QC  H2T 3E4 Canada.
%\end{bottomstuff}

%\maketitle

\section{Introduction}
%%%%%%%%%%%%%%%%%%%%%%%%%%%%%%%%%%%%%%%%%%%%%%%%%%%%%%%%%%%%%%%%%%%%%%%%%%%%%%%%%%%%%%%%%%%%%%%%%%
%%%%%%%%%%%%%%%%%%%%%%%%%%%%%%%%%%%%%%%%%%%%%%%%%%%%%%%%%%%%%%%%%%%%%%%%%%%%%%%%%%%%%%%%%%%%%%%%%%
\structure{The introduction motivates the problem without discussion any solution. }

In a data-driven world where permanent storage devices have an ever
growing capacity, I/O access becomes a bottleneck when read/write
performance does not increase as quickly as
capacity~\cite{vitter2002externalalgos}.  As Gray put it: ``we're able
to store more than we can access''~\cite{grayqueue}.  At the same
time, users of applications in OLAP~\cite{olap} or in
visualization~\cite{silva02outcore} now expect online processing of
their data sets. A strategy to solve this problem is to use a
relatively small internal-memory buffer to precompute some of the
expected queries.  For example, it might be reasonable to use a memory
buffer of one \mebibyte{} (1~\MiB{})\footnote{
Throughout, we  \ifthenelse{\boolean{evilunits}}
{use the familiar units of kB, MB, GB and TB to
%respectively
measure storage in
groups of $2^{10}$, $2^{20}$, $2^{30}$ and $2^{40}$ bytes.
Our units thus coincide with the new IEC
units KiB, MiB, GiB and TiB~\cite{iec:new-nonSI-units}.}
{use the preferred IEC units~\cite{iec:new-nonSI-units} of
\KiB{}, \MiB{}, \GiB{} and \TiB{}, which %respectively
measure storage in
groups of $2^{10}$, $2^{20}$, $2^{30}$, and $2^{40}$ bytes.}}
per \tebibyte{} (\TiB{}) of external-memory data.

If $X$ is a random variable with probability distribution $f$, we
define the moment of order $N$ about a value $c$
as the expectation of $(X-c)^N$ or
$E((X-c)^N)=\int (x-c)^N f(x) dx$.  Correspondingly, given an array $a$ and
some constant $c$, $\sum_i (i-c)^N a_i $ is a moment of order $N$. Given
an arbitrarily large array $a$, we can precompute the moments with
ease, however we are interested in \textbf{local} moments: given a
range of indices $p,\ldots,q$ and a constant $c$ all provided dynamically,
we wish to compute
$\sum_{i=p}^{q} (i-c)^N a_i $ online. Frequency moments~\cite{alon1996sca}, $\sum_i f^k_i$ where $f_i$ is the number of occurences of item $i$, are
outside the scope of this paper.

Local moments are used widely, from pattern recognition and image
processing to multidimensional databases (OLAP). Among other things,
they have been proposed as a replacement for maximum likelihood
methods in statistics, %in order
to cope with the magnitude of the new
data sets~\cite{binscott1997}.

As an example application, given the number of items in a store for
every possible price, the sum (moment~0) over a range of prices would
return how many items are available, the first moment would return the
total dollar value for items in the specified price range, and the
second moment would allow us to compute the standard deviation and variance
 of the
price.

As another example, imagine a moving sensor measuring the density of
some metallic compound in the ground.  A geophysicist could then ask
for the average density of this compound over some terrain or for the
``center of mass'' in some region. Indeed, if the array $a$ contains
the density over some one-dimensional strip of terrain, then the
average density over some region given by indices $p,\ldots,q$ is
given by $\sum_{i=p}^{q} a_i / (q-p)$ whereas the center of mass is
given by $({ \sum_{i=p}^{q} i a_i }) / { \sum_{i=p}^{q} a_i}$.

As yet another example, consider the local regression
problem~\cite{cleveland-smoothing}. Suppose that given a large segment $p,\ldots,q$
of a very large array $y_i$, the user wants to view the best
polynomial of order $N-1$ fitting the data:
this problem occurs when trying to  segment time series~\cite{lemire2007}, for example.
Given a polynomial
$\sum_{k=0}^{N-1} a_k x^k$, the residual energy is given by $\sum_{i=p}^{q}
(y_i - \sum_{k=0}^{N-1} a_k i^k)^2$.  Setting the derivative with
respect to $a_l$ to zero for $l=0,\ldots,N-1$, we get
a system of $N$ equations in $N$ unknowns,
$\sum_{k=0}^{N-1} a_k \sum_{i=p}^{q} i^{k+l}= \sum_{i=p}^{q} y_i i^l $
where $l=0,\ldots,N-1$.  Note that the right-hand-sides
of the equations are $N$ local moments whereas on the left-hand-side,
we have the $N \times N$ matrix
$A_{k,l}=\sum_{i=p}^{q} i^{k+l}$. For any $k+l$, we can compute
the sum $\sum_{i=p}^{q} i^{k+l}$ in constant time irrespective of $q-p$. To see
this, consider that because
 $\sum_{i=p}^{q} i^{k+l}= \sum_{i=0}^{q} i^{k+l}- \sum_{i=0}^{p-1} i^{k+l}$,
 it is sufficient to be able to compute expressions of the form
 $\sum_{i=0}^{q} i^{k+l}$ quickly.
However, there is %exists
 a  %general
 formula for these summations.

Several fast techniques have been proposed to compute local
moments~\cite{liimageseg,zhou95computing} but this paper is concerned
with \textbf{precomputing} auxiliary information
to speed up the computation of local moments.
%%%%%%%%%%%%%%%%%%%%%%%%%%%%%%%%%%%%%%%%%%%%%%%%%%%%%%%%%%%%%%%%%%%%%%%%%%%%%%%%%%%%%%%%%%%%%%%%%%
%%%%%%%%%%%%%%%%%%%%%%%%%%%%%%%%%%%%%%%%%%%%%%%%%%%%%%%%%%%%%%%%%%%%%%%%%%%%%%%%%%%%%%%%%%%%%%%%%%
\section{Notation}
%%%%%%%%%%%%%%%%%%%%%%%%%%%%%%%%%%%%%%%%%%%%%%%%%%%%%%%%%%%%%%%%%%%%%%%%%%%%%%%%%%%%%%%%%%%%%%%%%%
%%%%%%%%%%%%%%%%%%%%%%%%%%%%%%%%%%%%%%%%%%%%%%%%%%%%%%%%%%%%%%%%%%%%%%%%%%%%%%%%%%%%%%%%%%%%%%%%%%
\structure{A very mundane section. No real content. }

We use C-like indexing for arrays: an array of length $n$ is indexed
from $0$ to $n-1$ as in $a_0,\ldots,a_{n-1}$. Indices are always
integers, so that the notation $i\in[k,l]$ for an index means that $i
\in \{k,\ldots,l\}$.  Given a set $R$, the set of all finite arrays of
values in $R$ is denoted by $\mathcal{A}^R$.
The restriction of a function $f$ to domain $D$
is noted $f_{|D}$.
Some common functions
over real-valued arrays ( $\mathcal{A}^{\mathbb{R}}$) or ``range-query
functions'' include COUNT and SUM, where
$\mathrm{COUNT}(a_0,\ldots,a_{n-1}) = n$ and
$\mathrm{SUM}(a_0,\ldots,a_{n-1}) = \sum_{i=0}^n a_i$.  Other possible
queries include MAX (which returns the maximum value in a range) or
MAX\_N (which returns the N largest values found in a range).
The query moment~of~order~N is formally defined as
$\sum_{i=p}^{q} (i-p)^N a_i$.
Note that
computing query moments of order $N$, for various $N$,
allows the calculation of any local moment.
Indeed, notice that
\begin{equation}
\sum_{i=p}^{q} (i-c)^N a_i =\sum_{k=0}^{N}
{N\choose k}
(p-c)^{N-k} \sum_{i=p}^{q} (i-p)^k a_i \label{eqn:bin-expans}
\end{equation}
for any constant $c$,
by the expansion of $((i-p)+(p-c))^N$ using the Binomial Theorem.

%%%%%%%%%%%%%%%%%%%%%%%%%%%%%%%%%%%%%%%%%%%%%%%%%%%%%%%%%%%%%%%%%%%%%%%%%%%%%%%%%%%%%%%%%%%%%%%%%%
%%%%%%%%%%%%%%%%%%%%%%%%%%%%%%%%%%%%%%%%%%%%%%%%%%%%%%%%%%%%%%%%%%%%%%%%%%%%%%%%%%%%%%%%%%%%%%%%%%
\section{Related Work}
%%%%%%%%%%%%%%%%%%%%%%%%%%%%%%%%%%%%%%%%%%%%%%%%%%%%%%%%%%%%%%%%%%%%%%%%%%%%%%%%%%%%%%%%%%%%%%%%%%
%%%%%%%%%%%%%%%%%%%%%%%%%%%%%%%%%%%%%%%%%%%%%%%%%%%%%%%%%%%%%%%%%%%%%%%%%%%%%%%%%%%%%%%%%%%%%%%%%%

\structure{Here we do not mention bin buffering per se. This is just some related OLAP
techniques. In retrospect, it does not seem such an important section now. It could be moved later
in the paper.}

Using a buffer, we can compute
linear range queries in $O(1)$
time. For example, given an array $A=\{a_0,a_1,\ldots, a_{n-1}\}$,
we can use the Prefix Sum method~\cite{ho97range} and precompute
the array
$PS(A)=\{a_0,a_0+a_1,a_0+a_1+a_2,\ldots,a_0+\cdots+a_{n-1}\}$. By a mere
subtraction, we can then compute any range sum of the form
$a_k+a_{k+1}+\cdots+a_l$ since
\begin{align*}
a_k+\cdots+a_l = (a_0+\cdots+a_l) - (a_0+\cdots+a_{k-1}).
\end{align*}
 However, if a few data points
are updated, then all of $PS(A)$ may need to be recomputed:
updates require $O(n)$ time.
A more robust approach is the Relative Prefix Sum (RPS) method~\cite{geffner}
which buffers
the prefix sums only locally over blocks of size $b$.
For example,
\begin{align*}
RPS(A)=\{a_0,a_0+a_1,a_0+a_1+a_2,
a_3,a_3+a_4,\ldots\}
\end{align*}
when $b=3$. Clearly, $RPS(A)$ can be updated in time $O(b)$. To
 still achieve $O(1)$ query time, we use an \textit{overlay} buffer
\begin{align*}
a_0+a_1+a_2,a_0+\cdots+a_5,\ldots, a_0+\cdots+a_{n-1}
\end{align*}
that can be updated in time $O(n/b)$.
While RPS requires $\Theta(n)$ in storage, updates can be done in  $O(\sqrt{n})$
time by choosing $b=\sqrt{n}$. We can improve the
update performance of RPS using the Pyramidal Prefix Sum (PyRPS)
 method~\cite{cascon}.  Its query complexity is $O(\rho)$ with update
cost $O(\rho n^{1/\rho})$, where $\rho =2,3,\ldots$  Thus, PyRPS obtains constant-time
queries but with faster updates than RPS\@.
Furthermore, PyRPS supports queries and updates in logarithmic time by choosing $\rho=\log(n)$.

Of course, these techniques extend to other range queries.
In the context of orthogonal range queries for multidimensional databases,
similar results are even possible for range-maximum
queries~\cite{poonortho2003}.

The PS, RPS, and PyRPS methods (and similar alternatives for other
queries) have a common inconvenience: each
type of range query is buffered separately: $\sum_k a_k$, $\sum_k k
a_k$, $\sum_k k^2 a_k$, $\ldots$ An equivalent view is of a single
\emph{tuple-valued} buffer, rather than several real-valued buffers.
The ProPolyne framework~\cite{propolyne} showed how to avoid tuple-valued
buffers by using wavelets. ProPolyne has both logarithmic
queries and updates at the cost of a buffer of size $O(n)$ for
computing local moments (\textit{Polynomial Range Queries}).
ProPolyne simultaneously buffers all local moments up to a given
degree.  So, ProPolyne reduces storage when compared with
prefix-sum methods such as PyRPS, but at the expense of constant-time
queries.
See Table~\ref{cap:Comparison-of-a} for a comparison of
various alternatives to buffer local moments.

%We should note that i
In a wavelet framework such as ProPolyne,
we can keep only the most
significant wavelet coefficients to reduce storage and increase performance,
while obtaining reasonably accurate
results~\cite{Approximatewavelets,1066189,Vitterdatacubeapprox}.
It is also possible to process the queries incrementally so that approximate results
are available sooner.

\ifthenelse{\boolean{acmformat}}{
\begin{acmtable}{0.87\columnwidth}
\centering
\renewcommand{\arraystretch}{1.2}
\begin{small}
\begin{tabular}{|p{0.365\columnwidth}|c|c|c|}
Algorithm&
Query&
Update&
Storage\tabularnewline
\hline
\textsc{One-Scale Ola}&
\textbf{$O(N n / b + N^2b)$} &
\textbf{$O(N)$}&
\textbf{$n/b+1$}\tabularnewline
\textsc{Hierarchical Ola}&
\textbf{$O(N^2 b \log_b n)$}&
\textbf{$O(N^2 \log_b n)$}&
\textbf{$n/b+1$}\tabularnewline
Bin Buffering&
$O( N n / b + Nb)$ &
$O(N)$&
$N n/b$\tabularnewline
Hierarchical Bin Buffering&
$O(N b \log_b{n})$ &
$O(N  \log_b n)$&
$N n/b$\tabularnewline
ProPolyne&
$O(N^2 \log_2 n)$&
$O(N^2 \log_2 n)$ &
$n$\tabularnewline
Prefix Sums&
$O(N )$&
$O(N n)$&
$N n$\tabularnewline
Relative Prefix&
$O(N )$&
$O(N \sqrt{n})$&
$N n$\tabularnewline
PyRPS&
$O(N \rho )$&
$O(N \rho \sqrt[\rho]{n})$ &
$N n$\tabularnewline
PyRPS (log)&
$O(N \log n)$&
$O(N \log n)$&
$N n$\tabularnewline
\end{tabular}\end{small}
\caption{\label{cap:Comparison-of-a}Comparison of local
moment algorithms with corresponding storage requirements
and complexity for large $n$ where we buffer the first $N$ moments.
Note that $\rho=2,3\ldots$ is a parameter that  can be chosen to be
large. The storage requirement is %in terms of
the number of components
needed to buffer computations. \textsc{Ola} is a form of
Bin Buffering specific to local moments.
}
\end{acmtable}
}{
\begin{table}
\caption{\label{cap:Comparison-of-a}Comparison of local
moment algorithms with corresponding storage requirements
and complexity for large $n$ where we buffer the first $N$ moments.
Note that $\rho=2,3\ldots$ is a parameter that  can be chosen to be
large. The storage requirement is %in terms of
the number of components
needed to buffer computations. \textsc{Ola} is a form of
Bin Buffering specific to local moments.
}
\renewcommand{\arraystretch}{1.2}
\begin{small}
\begin{center} \begin{tabular}{|p{0.2\columnwidth}|c|c|c|}
\hline
Algorithm&
Query&
Update&
Storage\tabularnewline
\hline
\hline
\textsc{One-Scale Ola}&
\textbf{$O(N n / b + N^2b)$} &
\textbf{$O(N)$}&
\textbf{$n/b+1$}\tabularnewline
\hline
\textsc{Hierarchical Ola}&
\textbf{$O(N^2 b \log_b n)$}&
\textbf{$O(N^2 \log_b n)$}&
\textbf{$n/b+1$}\tabularnewline
\hline
Bin Buffering&
$O( N n / b + Nb)$ &
$O(N)$&
$N n/b$\tabularnewline
\hline
Hierarchical Bin Buffering&
$O(N b \log_b{n})$ &
$O(N  \log_b n)$&
$N n/b$\tabularnewline
\hline
ProPolyne&
$O(N^2 \log_2 n)$&
$O(N^2 \log_2 n)$ &
$n$\tabularnewline
\hline
Prefix Sums&
$O(N )$&
$O(N n)$&
$N n$\tabularnewline
\hline
Relative Prefix&
$O(N )$&
$O(N \sqrt{n})$&
$N n$\tabularnewline
\hline
PyRPS&
$O(N \rho )$&
$O(N \rho \sqrt[\rho]{n})$ &
$N n$\tabularnewline
\hline
PyRPS (log)&
$O(N \log n)$&
$O(N \log n)$&
$N n$\tabularnewline
\hline
\end{tabular}\end{center}\end{small}
\end{table}
}

%%%%%%%%%%%%%%%%%%%%%%%%%%%%%%%%%%%%%%%%%%%%%%%%%%%%%%%%%%%%%%%%%%%%%%%%%%%%%%%%%%%%%%%%%%%%%%%%%%
%%%%%%%%%%%%%%%%%%%%%%%%%%%%%%%%%%%%%%%%%%%%%%%%%%%%%%%%%%%%%%%%%%%%%%%%%%%%%%%%%%%%%%%%%%%%%%%%%%
\section{Contribution and Organization}
%%%%%%%%%%%%%%%%%%%%%%%%%%%%%%%%%%%%%%%%%%%%%%%%%%%%%%%%%%%%%%%%%%%%%%%%%%%%%%%%%%%%%%%%%%%%%%%%%%
%%%%%%%%%%%%%%%%%%%%%%%%%%%%%%%%%%%%%%%%%%%%%%%%%%%%%%%%%%%%%%%%%%%%%%%%%%%%%%%%%%%%%%%%%%%%%%%%%%

\structure{This is where we sell the paper. }

For storage, a reduction from $n$ to $\sqrt{n}$ can be
quite significant: if $n=2^{40}$ (1~\TiB{}), then $\sqrt{n}=2^{20}$ (1~\MiB{}),
so we argue that simple buffering schemes
might often prove
more practical than PyRPS or ProPolyne, especially
because a small buffer can be generally constructed faster.
As Ho et al.~\cite{ho97range} observed (for the Prefix Sum Method),
good performance can be obtained with a small buffer,
provided we retain access to the original array.

The paper is organized as follows. We first
consider how bin buffers
can be used to speed up
many range queries over dynamic external
arrays (section~\ref{sec:bb}). For each bin or ``range of indices,''
Bin Buffering~\cite{moerkotte98} associates a
single buffer component.
However, its scalability is limited because very large buffers do not improve
performance and can even worsen it. In section~\ref{sec:hbb},  we present
the analysis of a variant,
Hierarchical Bin Buffering, which supports logarithmic queries and updates
even for modest buffers. These results are novel, but have been alluded
to in the concluding section of a paper~\cite{moerkotte98}.
In section~\ref{sec:obb}, we present a novel buffering framework: Overlapped Bin Buffering.
In Overlapped Bin Buffering, each buffer component depends not only on one but a range
of bins. In this context, we present Lagrange Interpolation (section~\ref{sec:li}) as a tool
to compute a useful buffer and establish an explicit link between buffering and interpolation.
The result is an \textsc{Ola} buffer and we show it can be hierarchical as well.
Section~\ref{sec:exp} presents precisely stated algorithms and experimental results.
We then proceed on some concluding remarks. We elaborate on the
differences between \textsc{Ola} and \textsc{Bin Buffering} (section~\ref{sec:obbvshbb})
and show how \textsc{Ola} can support efficient progressive approximate queries (section~\ref{sec:approxobb}).

%%%%%%%%%%%%%%%%%%%%%%%%%%%%%%%%%%%%%%%%%%%%%%%%%%%%%%%%%%%%%%%%%%%%%%%%%%%%%%%%%%%%%%%%%%%%%%%%%%
%%%%%%%%%%%%%%%%%%%%%%%%%%%%%%%%%%%%%%%%%%%%%%%%%%%%%%%%%%%%%%%%%%%%%%%%%%%%%%%%%%%%%%%%%%%%%%%%%%
\section{Fast Algebraic Range Queries using Precomputation over Bins}
\label{sec:bb}
%%%%%%%%%%%%%%%%%%%%%%%%%%%%%%%%%%%%%%%%%%%%%%%%%%%%%%%%%%%%%%%%%%%%%%%%%%%%%%%%%%%%%%%%%%%%%%%%%%
%%%%%%%%%%%%%%%%%%%%%%%%%%%%%%%%%%%%%%%%%%%%%%%%%%%%%%%%%%%%%%%%%%%%%%%%%%%%%%%%%%%%%%%%%%%%%%%%%%

\structure{This is the somewhat theoretical high level part of the paper where we lay a
theoretical foundation which we only partly used in the rest of the paper. It is fairly
close in content to Moerkotte, but from a different point of view.}

Let $R$ be an algebraic structure such as $\mathbb{R}$ or $\mathbb{R}^m$.
A range-query function $Q:\mathcal{A}^{R} \rightarrow R$ is
\emph{distributive}~\cite{gray:datacube} if there is a function
$F:\mathcal{A}^{R} \rightarrow R$ such that
for all $0 \leq k < n-1$,
\[
 Q(a_0,\ldots, a_k,a_{k+1},\ldots,a_{n-1})  =  F( Q(a_0,\ldots, a_k),
Q(a_{k+1},\ldots,a_{n-1})).
\]

Examples of distributive range-query functions include
COUNT, SUM, and MAX\@. In this paper, we shall only consider range
queries where the computational cost is independent of the values
being aggregated.
 By convention,  $F(Q(a_0,\ldots, a_k)) =  Q(a_0,\ldots, a_k)$; i.e.,
the function $F$ is the identity when applied to a single value.

We have
\begin{eqnarray*}
Q(a_0,\ldots,a_{n-1})& =& F(Q(a_0,\ldots,a_{n-1})) \\
& = &F(Q(a_0,\ldots,a_{n-2}), Q(a_{n-1})) \\
& = &F(F(Q(a_0,\ldots,a_{n-2})),Q(a_{n-1})) \\
& =& F( F( F( Q(a_0,\ldots,a_{n-3}),Q(a_{n-2})), \\
&& Q(a_{n-1}))\\
& =& \cdots
\end{eqnarray*}
and so we can compute
$F(Q(a_0,\ldots,a_{n-1}))$
recursively, using $n-1$
pairwise aggregations. Hence, a distributive range query over $n$~terms has complexity $O(n)$.

Distributive range-query functions can be combined. For example,
define the joint query function $(Q_1,Q_2)$
as the tuple-valued query function
\[
(Q_1,Q_2)(a_0,\ldots,a_{n-1}) =(Q_1(a_0,\ldots,a_{n-1}),
Q_2(a_0,\ldots,a_{n-1})).
\]
We can verify that $(Q_1,Q_2)$ is distributive if $Q_1$ and $Q_2$ are
distributive.

In this paper, %we say that
a real-valued range-query
function $Q:\mathcal{A}^{\mathbb{R}} \rightarrow \mathbb{R}$ is
\emph{algebraic} if there is an intermediate tuple-valued
\textbf{distributive} range-query function $G:\mathcal{A}^{\mathbb{R}}
\rightarrow \mathbb{R}^m$ from which $Q$ can be computed.
 For example, given the tuple $(\mathrm{COUNT},\mathrm{SUM})$, one
can compute AVERAGE by a mere ratio, so AVERAGE is an example of an
algebraic query function.  In other words, if $Q$ is an algebraic
function then there must exist $G$ and $F:\mathcal{A}^{\mathbb{R}^m}
\rightarrow \mathbb{R}^m$ for some fixed integer $m$ such that
 \[
 G(a_0,\ldots, a_k,a_{k+1},\ldots,a_{n-1})  = F( G(a_0,\ldots, a_k),
                                                  G( a_{k+1},\ldots,a_{n-1})).
\]
We have that algebraic queries can be computed in time $O(n)$ because we
can interpret them as distributive queries over tuples.
All real-valued distributive range-query functions are algebraic, but
examples of non-distributive algebraic functions
include MAX\_N, AVERAGE, CENTER\_OF\_MASS, STANDARD\_DEVIATION, and local moments.
%in general.
As an example, if $Q$ is AVERAGE, then we can choose $G$ to
compute the tuples (COUNT,~SUM) and $F$ can be the component-wise sum.
Similarly, if $Q$ is the local moment
of order $N$, then $G$ should compute
the tuple made of (COUNT,~SUM,~$\ldots$,~moment of order~N).
%In general, observe that c
Combining two $N^{th}$-order
query moments $\sum_{i=p}^{r-1}(i-p)^N a_i$
and $\sum_{i=r}^q(i-r)^N a_i$ into an aggregate $\sum_{i=p}^q(i-p)^N a_i$ can be
done by Eq.~(\ref{eqn:bin-expans}), which requires that we
know the lower-order moments.
In terms of $N$, a straightforward implementation of $G$
runs in O($N$) time. Indeed, we can compute $G(a_0,\ldots,a_k)$ as
$F(G(a_0,\ldots,a_{k-1}),G(a_k))$ but $G(a_k)= (1,a_k,0,\ldots,0)$.
Operation $F$ itself requires $O(N)$ time. Hence,
$G(a_0,\ldots,a_k)$ can be computed in $O(Nn)$ time.

In what follows, we will specify algebraic functions as
triples $(Q,G,F)$. Whenever
we use the $(Q,G,F)$ notation, it is understood that
\[
Q:\mathcal{A}^{\mathbb{R}} \rightarrow \mathbb{R},\mbox{\ and\  }
G:\mathcal{A}^{\mathbb{R}} \rightarrow \mathbb{R}^m, \mbox{\ and\ }
F:\mathcal{A}^{\mathbb{R}^m} \rightarrow \mathbb{R}^m
\]
and often $Q$ will not be used explicitly because computing $G$ is enough.
We will assume that the size of the tuples, $m$, is small:  $m \lesssim 16$.

\begin{figure}
\begin{center}
\includegraphics[width = 0.8\columnwidth, angle = 0]{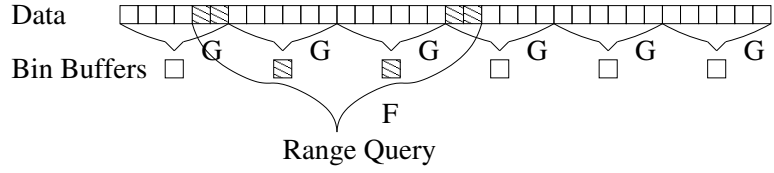}
\end{center}
\caption{\label{fig:binbuffering}An algebraic range query supported
by Bin Buffering, as in Eq.~(\ref{eqn:firstlevelbin}).}
\end{figure}

Given an integer $b$ that divides $n$
and given an algebraic
function $(Q,G,F)$, we can buffer queries by precomputing $b/n$ components
\begin{align*}
	& G(a_0,\ldots,a_{b-1}),\\
	& G(a_{b},\ldots,a_{2b-1}),\ldots,\\
	& G(a_{n-b},\ldots,a_{n-1})
\end{align*}
denoted $B_0,\ldots,B_{n/b-1}$.
This buffer can be updated in time $O(b)$ if
an array component is changed.
Using this precomputed array, range queries can be
computed in time $O(n/b+b)$ because of the formula
 \begin{equation}\label{eqn:firstlevelbin}
 	G(a_k,\ldots,a_l)  =
        F(G(a_k, \ldots, a_{b \lceil k/b \rceil - 1}),
        B_{\lceil k/b \rceil}, \ldots,
        B_{\lfloor l/b \rfloor - 1},
	G(a_{b \lfloor l/b \rfloor }, \ldots, a_{l})).
  \end{equation}
  See also Figure~\ref{fig:binbuffering}.
By choosing $b=\sqrt{n}$, we get updates
and queries in time $O(\sqrt{n})$ with a buffer
of size $\sqrt{n}$. In different terms, this algorithm was
presented by Moerkotte~\cite{moerkotte98}.

When buffering local moments of order $N$, $G$ computes
$N+1$-tuples
so
that the size of the buffer is $(N+1) \times n/b$. This can be
reduced to $N \times n/b$ if all bins are of a fixed size $b$, since
we need not store COUNT.

An algebraic range-query function $(Q,G,F)$ is \textit{linear} if the corresponding intermediate query $G$
satisfies
\[
G(a_0+\alpha d_0,\ldots,a_{n-1}+\alpha d_{n-1})= G(a_0,\ldots,a_{n-1})
+ \alpha  G(d_0,\ldots,d_{n-1})
\]
for all arrays $a,d$, and constants $\alpha$.
SUM, AVERAGE and local moments are linear functions; %observe that
MAX is not linear.
Linear queries over bins of size $b$ can be computed using the formula
$G(a_0,\ldots,a_{b-1}) = a_0 G(e^{(0)})+\ldots+a_{b-1} G(e^{(b-1)})$,
where $e^{(i)}$ an array of size $b$ satisfying $e^{(i)}_j = 0$ if $i\neq j$ and $e^{(i)}_i = 1$.
For our purposes, we define an update by the location of the change, $k$,
and by how much the value changed, $\Delta=a'_k - a_k$.
We see that the update complexity for buffered linear range queries is reduced to constant time since
\begin{align*}
G(a_0,\ldots,a_{k-1},a'_k,a_{k+1},\ldots,a_{b-1})
- G(a_0,\ldots,a_k,\ldots,a_{b-1}) &= (a'_k - a_k) G(e^{(k)})\\
 & = G(e^{(k)}) \Delta
\end{align*}
and $G(e^{(k)})$ can be precomputed or computed in constant time.

Hence, we see that:
\begin{enumerate}
\item All algebraic queries can be bin buffered, including MAX, AVERAGE, and local moments.
\item For linear queries, the buffer can be updated quickly.
\end{enumerate}

\begin{lemma}For an algebraic range-query function $Q$,
given an array of size $n$,
Bin Buffering uses a buffer of $n/b$ tuples computed
in time $O(n)$ and updated in time  $O(b)$ to support
queries in time $O(n/b+b)$.
Choosing $b=\sqrt{n}$ minimizes the query complexity
to $O(\sqrt{n})$.
If $Q$ is linear then updates take constant time.
\end{lemma}

\begin{lemma}Consider  local moments of degree $N$, where $N$ is fixed and small.
Given an array of size $n$,
Bin Buffering uses a buffer of size $N \times n/b$ computed
in time $O(n)$ and updated in constant time  to support
queries in time $O(n/b+b)$.
Choosing $b=\sqrt{n}$ minimizes the query complexity
to $O(\sqrt{n})$.
\end{lemma}

If $N$ is not considered fixed, buffer computation is $O(Nn)$,
update time is $O(N)$ and query time is  $O(N n/b + b N )$.

A possible drawback of the Bin Buffering algorithm is that the query
complexity cannot be reduced by using a larger buffer.
For example, in going from a buffer of size
$n/2$ to a buffer of size $\sqrt{n}$, the algorithm's complexity %actually
goes down from
$O(n)$ to $O(\sqrt{n})$. We will show in the next
section how we can use larger buffers in a hierarchical setting
to increase the speed.

%%%%%%%%%%%%%%%%%%%%%%%%%%%%%%%%%%%%%%%%%%%%%%%%%%%%%%%%%%%%%%%%%%%%%%%%%%%%%%%%%%%%%%%%%%%%%%%%%%
%%%%%%%%%%%%%%%%%%%%%%%%%%%%%%%%%%%%%%%%%%%%%%%%%%%%%%%%%%%%%%%%%%%%%%%%%%%%%%%%%%%%%%%%%%%%%%%%%%
\section{Hierarchical Bin Buffering}
\label{sec:hbb}
%%%%%%%%%%%%%%%%%%%%%%%%%%%%%%%%%%%%%%%%%%%%%%%%%%%%%%%%%%%%%%%%%%%%%%%%%%%%%%%%%%%%%%%%%%%%%%%%%%
%%%%%%%%%%%%%%%%%%%%%%%%%%%%%%%%%%%%%%%%%%%%%%%%%%%%%%%%%%%%%%%%%%%%%%%%%%%%%%%%%%%%%%%%%%%%%%%%%%

\structure{This is where we elaborate more on Moerkotte's paragraph.}

In the previous section, we showed we could
precompute algebraic range queries over bins of size $b$
%in order
 to support $O(n/b+b)$-time queries. We can scale this
 up using a pyramidal or hierarchical approach~\cite{cascon}.

\begin{figure*}
\begin{center}
\includegraphics[width = 12cm, angle = 0]{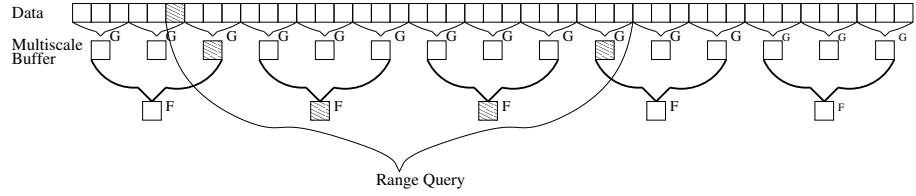}
\end{center}
\caption{\label{fig:multiscalebinbuffering}An algebraic range query supported
by Hierarchical Bin Buffering. We essentially repeat
Figure~\ref{fig:binbuffering} over the buffer itself.
 In the example
given, by aggregating buffer components, we replace 6 first-scale buffer
components by 2 second-scale components. }
\end{figure*}

For a fixed $b$, the $n/b$ term dominates the $O(n/b+b)$
complexity.  %Observe that i
In Eq.~(\ref{eqn:firstlevelbin}) the
$n/b$ term comes from the buffer aggregation.  So, we started from an
aggregation over $n$~terms and reduced it to an aggregation over $n/b$~terms; clearly we can further reduce the aggregation over $n/b$~terms
to an aggregation over $n/b^2$~terms by the same technique (see
Figure~\ref{fig:multiscalebinbuffering}).  In other words, we can buffer the
buffer.  Hence, considering the buffer of size $n/b$ as a source
array, we can buffer it using $n/b^2$ components %in order
to support
queries in time $O(n/b^2+b)$ over the buffer instead of $O(n/b)$.
Thus, the end result is to have queries in time $O(n/b^2+2b)$ with a
buffer of size at most $n/b+n/b^2$. Repeating this argument $\log_b n$
times, we get queries in time $O(b \log_b n)$ using
$\sum_{k=1,\ldots,\log_b n} n/b^k \leq n/(b-1)$ storage.
If the query function
is \emph{invertible}, as defined %shortly,
in the next subsection,
then
we can use
in-place storage for higher-scale buffers.  This
reduces the internal memory
usage to $n/b$.  The update complexity is $O(b \log_b n)$ in general
and $O( \log_b n)$ for linear queries.
%%%%%%%%%%%%%%%%%%%%%%%%%%%%%%%%%%%%%%%%%%%%%%%%%%%%%%%%%%%%%%%%%%%%%%%%%%%%%%%%%%%%%%%%%%%%%%%%%%
%%%%%%%%%%%%%%%%%%%%%%%%%%%%%%%%%%%%%%%%%%%%%%%%%%%%%%%%%%%%%%%%%%%%%%%%%%%%%%%%%%%%%%%%%%%%%%%%%%
\subsection{In-place Storage for Invertible Query Functions}
%%%%%%%%%%%%%%%%%%%%%%%%%%%%%%%%%%%%%%%%%%%%%%%%%%%%%%%%%%%%%%%%%%%%%%%%%%%%%%%%%%%%%%%%%%%%%%%%%%
%%%%%%%%%%%%%%%%%%%%%%%%%%%%%%%%%%%%%%%%%%%%%%%%%%%%%%%%%%%%%%%%%%%%%%%%%%%%%%%%%%%%%%%%%%%%%%%%%%
A given algebraic
%Given  an algebraic
function $(Q,G,F)$, %we say
%it :  owen killed the ``it'' because it seemed awkward
is \emph{invertible} if O(1) time is
sufficient to solve for
$x$ in $z=F(x,y)$, where $x,y,z \in \mathbb{R}^m $
($m$ is assumed small).
%Note that l
Linear queries are invertible, and %B   --- owen felt the small sentence choppy
% although the resulting sentence is a tad complex.
being invertible is a useful property: it means that the storage used by
$x$ can be used to store $z$ --- storing $x,y$ or $z,y$
is almost equivalent.
This lets us ``buffer a buffer'' in place, as the
next proposition shows.

\begin{proposition}\label{invertibleischeap}If $(Q,G,F)$ is an invertible
algebraic query function,
then
% word shifting to fix line break problem
% Owen: reverting for TALG...
%%we can store
%
% Owen wonders if the hyphen below is still making an overfull line?
% Daniel thinks that the "the" should be removed, this fixed the overfull
 second-scale Bin Buffer
components $B'_0=F(B_0,\ldots,B_{b-1})$ and $B'_b=F(B_b,\ldots,B_{2b-1}),$ \ldots
can be stored  %% back it goes for TALG
in-place at positions $0,b,2b,\ldots$ in the buffer
(overwriting values $B_0, B_b, B_{2b},\ldots$) without increasing query time complexity.
\end{proposition}
\begin{proof}
Assume we use in-place storage for the second-scale Bin Buffers
$B'_0, B'_1,\ldots$, overwriting $B_0,B_b,\ldots$.
We
must
evaluate expressions of the form $F(B_{k}, \ldots,B_{l})$,
which can be done using the second-scale Bin Buffer,
according to the formula
 \begin{align*}
 	& F(B_k,\ldots,B_l)
        =
        F(F(B_k, \ldots, B_{\lceil k/b \rceil b - 1}),
        B'_{b \lceil k/b \rceil}, \ldots,
        B'_{b(\lfloor l/b \rfloor - 1)},
	F(B_{b \lfloor l/b \rfloor }, \ldots, B_{l})).
  \end{align*}
The only place where an overwritten value appears is in the last
term: $B_{b \lfloor l/b \rfloor }$ has been replaced by the value of
     $B'_{b \lfloor l/b \rfloor}$.
However, the query is invertible, so $B_{b \lfloor l/b \rfloor }$ can be
recovered in constant time.
Thus the algorithm using two-scale buffers is still going to be $O(n/b^2+2b)$,
even though in-place storage has been used.
\end{proof}

We can repeat this process for each buffer scale, each time incurring only
a fixed cost for recovering an overwritten value.
The total additional cost is $O(\log_b n)$, but this is dominated by
the cost of the query itself ($O(b\log_b n)$). In other words,
in-place storage almost comes for free.

As an example of Hierarchical Bin Buffering with in-place storage,
consider the array $a_0,\ldots,a_{80}$ ($n=81$)
and some invertible algebraic query function $(Q,G,F)$. The one-scale
Bin Buffering algorithm with $b=3$ simply precomputes
\begin{align*}
B_0=G(a_0,a_1,a_2),
B_1=G(a_3,a_4,a_5),\ldots,
B_{26}=G(a_{78},a_{79},a_{80})
\end{align*}
so that if we want $Q(a_1,\ldots,a_{79})$, we still
have to compute
\begin{align*}F(G(a_1,a_2),B_1,\ldots,B_{25},G(a_{78},a_{79}))\end{align*}
which is the aggregation of 27~terms using $F$.
We can aggregate the buffer itself in a second buffer, in this case,
by precomputing
\begin{align*}
 B'_0=F(B_0,B_1,B_2),B'_3=F(B_3,B_4,B_5),\ldots,
 B'_{24}=F(B_6,B_7,B_8)
 \end{align*}
and storing them in-place, so that $B_0,B_3,\ldots,B_{24}$ are replaced by
the newly computed $B'_0,B'_3,\ldots,B'_{24}$.
Then, to compute $Q(a_1,\ldots,a_{79})$, it suffices to compute
\begin{align*}F(G(a_1,a_2),B_1,B_2,B'_3,B'_6,\ldots,B'_{21},B_{24},B_{25},G(a_{78},a_{79})),\end{align*}
the aggregation of only
13~terms. The query cost is halved without using any
additional memory.

As the next lemma explains, the hierarchical case presented above
is simply a generalization of the case in the previous section, but where
large buffers can be used to answer queries in logarithmic time.
Recall that local moments are linear and invertible whereas MAX queries
are neither.

\begin{lemma}\label{lem:analysis-algeb-hbb}
For an algebraic range-query function $Q$,
given an array of size $n$,
Hierarchical Bin Buffering uses a buffer of $\frac{n}{b-1}$ tuples computed
in time $O(n)$ and updated in time  $O(b \log_b n)$ to support
queries in logarithmic time $O(b \log_b n)$.
If the query is invertible, then a smaller memory
buffer of size $n/b$
can be used; for linear queries updates can be done in time $O(\log_b n)$.
\end{lemma}

For non-invertible queries such as MAX, that is, the worst
case scenario, this last lemma implies
that a buffer of size $n/(b-1)$ can support queries in time $O(b \log_b n)$
with updates in time $O(b \log_b n)$. For invertible and linear queries
such as SUM, the storage is only $n/b$ with updates in time $O( \log_b n)$.
Choosing $b=2$ minimizes the query complexity ($O(\log_2 n)$)
while maximizing the storage requirement at $n/2$, whereas choosing
$b=\sqrt{n}$ reduces to the non-hierarchical (one-scale) case with a query
complexity of $O(\sqrt{n})$ and a storage requirement of $\sqrt{n}$.

Note that
$G$ operates on tuples, and thus a buffer of $n/b$ elements %actually
occupies $mn/b$ space,  offsetting the economical
nature of the Hierarchical Bin Buffering algorithm.
Another result is that the $G$ operation becomes more expensive
for higher-order moments; in the analysis leading up to
Lemma~\ref{lem:analysis-algeb-hbb}, we implicitly assumed
the cost of $G$ was constant.  However, if the analysis considers
that operation costs increase with $N$, we have that
buffer construction is in $O(Nn)$, updates are in
$O(N \log_b n)$
and queries are in
$O(N b \log_b n)$.
As we shall see in the next
section, for some types of queries such as local moments, it is
possible to avoid using tuples in the buffer.

%%%%%%%%%%%%%%%%%%%%%%%%%%%%%%%%%%%%%%%%%%%%%%%%%%%%%%%%%%%%%%%%%%%%%%%%%%%%%%%%%%%%%%%%%%%%%%%%%%
%%%%%%%%%%%%%%%%%%%%%%%%%%%%%%%%%%%%%%%%%%%%%%%%%%%%%%%%%%%%%%%%%%%%%%%%%%%%%%%%%%%%%%%%%%%%%%%%%%
\section{Overlapped Bin Buffering}
\label{sec:obb}
%%%%%%%%%%%%%%%%%%%%%%%%%%%%%%%%%%%%%%%%%%%%%%%%%%%%%%%%%%%%%%%%%%%%%%%%%%%%%%%%%%%%%%%%%%%%%%%%%%
%%%%%%%%%%%%%%%%%%%%%%%%%%%%%%%%%%%%%%%%%%%%%%%%%%%%%%%%%%%%%%%%%%%%%%%%%%%%%%%%%%%%%%%%%%%%%%%%%%

In the previous sections, we described Bin Buffering and Hierarchical
Bin Buffering as it applies to all algebraic queries.
Such
Bin Buffering
is characterized by the facts that buffer components, $B_0=G(a_0,\ldots,a_{b-1})$,
$B_1=G(a_{b},\ldots,a_{2b-1})$,~$\ldots$, are over disjoint bins
and are aggregated using $G$ itself. In this
section, we will
consider only weighted sums as aggregate operators, and we will
define buffer components that depend on several bins at once.
This can also be interpreted as having overlapping bins.
Our motivation
is to buffer local moments using a single real-valued buffer, and
we begin by considering one-scale buffering in
subsections~\ref{section:generalcase}--\ref{section:piecewise},
but in subsection~\ref{section:hierarchicalOLA} we will extend
our results to the hierarchical case.

\subsection{General Case}
\label{section:generalcase}

Consider an array $a$ of size $n$ indexed as $a_0,\ldots,a_{n-1}$.
By convention, $a_j=0$ for $j \notin [0,n)$.
Assuming that $n$ is divisible by $b$, we group the terms in bins
of size $b$:  first $a_0,\ldots, a_{b-1}$, then $a_b,\ldots, a_{2b-1}$ and so on.
We have $n/b$ bins and we want to compute $n/b + 1$ buffer components
$B_0,\ldots,B_{n/b}$ %in order
to speed up some range-query functions
such as SUM\@. However, we drop the requirement that each buffer component
correspond to one and only one bin, but rather, we allow buffer components
to depend on several bins, hence the term ``Overlapped Bin Buffering.''

\begin{figure}
\begin{center}
\includegraphics[width = 0.8\columnwidth, angle = 0]{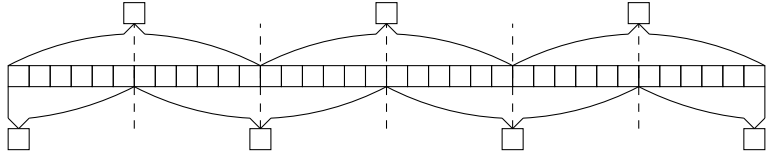}
\end{center}
\caption{\label{fig:overlapbinbuffering}Overlapped Bin Buffering with
$M=1,M'=1$. Buffered components depend on overlapping areas.}
\end{figure}

For an array $a_0,\ldots, a_{n-1}$ and given integers $M,M'\geq0$,
consider buffer components of the form
\begin{align*}B_k = \sum_{j=-Mb}^{M'b-1} c_{j} a_{j+kb}\end{align*}
where the coefficients $c_{j}$ are to be determined but are
zero outside their range, i.e.,
$c_{j}=0$ if $j \notin [-Mb,M'b)$. We could %easily
generalize this framework
so that $M+M'$ remains a constant but that $M$ and $M'$ depend on the bins:
we could accommodate the end and the beginning of the array so that $j+kb$ is always
in $[0,n)$, but this makes the formulas and algorithms more pedantic.
In essence, the $B_k$ are weighted sums over a range of $M+M'$ bins.
We can
interpret the coefficients $c_{j}$ as weights used to compute the buffer
component $B_k$, where $j$ gives the offset in the original array with
respect to $kb$.
An example is shown in Figure~\ref{fig:overlapbinbuffering}.  Note that
in the special case where $M=0, M'=1$, we have the usual Bin Buffering
approach, which maps each bin to exactly one buffer component
and where $c_j=1$ for $j\in [0,b)$.
As we shall see, increasing $M$ and $M'$ allows for additional degrees
of freedom.

%Notice that w
We can compute $\sum_{i=p}^{q} f(i) a_i$, by replacing $f$
by $\tilde{f}$ such that $\tilde{f}$ agrees with $f$ on $[p,q]$ but is
zero otherwise, and then compute $\sum_{i=0}^{n-1} \tilde{f}(i)
a_i$. Thus, it is enough to have a fast algorithm to compute
$\sum_{i=0}^{n-1} f(i) a_i$ for an arbitrary $f$.
The next proposition presents a formula that is instrumental in
achieving a fast algorithm.

\begin{proposition}\label{propform} Given an array $a_0,\ldots,a_{n-1}$ and integers
$M,M' \geq 0$,  and given the
buffer components $B_k = \sum_{j=-Mb}^{M'b-1} c_{j} a_{j+kb}$,
we have
\begin{align*}
\sum_{i=0}^{n-1} f(i) a_i
 = \sum_{k=0}^{n/b} f(kb) B_k
+ \sum_{i=0}^{n-1} \delta (i) a_i\label{eq:bignasty}
\end{align*}
where
\begin{align*}
\delta (i) =
f(i) -\sum_{k=\lfloor \frac{i}{b} \rfloor - M' + 1}^{\lfloor \frac{i}{b} \rfloor + M}  f(kb)  c_{i-kb}.
\end{align*}
\end{proposition}
\begin{proof}
By the definition of $B_k$, we have
\begin{align*}
\sum_{k=0}^{n/b} f(kb) B_k  =  \sum_{k=0}^{n/b} \sum_{j=-Mb}^{M'b-1} f(kb)  c_{j} a_{j+kb}.
\end{align*}
Define $i= j+kb$ so that
\begin{align*}
\sum_{k=0}^{n/b} \sum_{j=-Mb}^{M'b-1} f(kb) c_{j} a_{j+kb} =
\sum_{k=0}^{n/b} \sum_{i=kb-Mb}^{kb+M'b-1} f(kb)  c_{i-kb} a_{i}.
\end{align*}
Because $c_{i-kb}$ is zero whenever $i \notin [kb-Mb,kb+M'b)$, we can
replace $\sum_{i=kb-Mb}^{kb+M'b-1}$ by $\sum_{i=0}^{n-1}$ in the above equation
to get, after permuting the sums,
\begin{align*}
\sum_{k=0}^{n/b} f(kb) B_k
 =  \sum_{i=0}^{n-1} \left ( \sum_{k=0}^{n/b} f(kb)  c_{i-kb}  \right ) a_{i}.
\end{align*}
However, note that $c_{i-kb}$ is zero whenever $kb \notin (i-M'b,i+Mb]$ or
$k \notin  (  \lfloor \frac{i}{b} \rfloor - M', \lfloor \frac{i}{b} \rfloor + M]$. Hence, we can replace
$\sum_{k=0}^{n/b}$ by $\sum_{k=\lfloor \frac{i}{b} \rfloor - M' + 1}^{\lfloor \frac{i}{b} \rfloor + M} $
to get
\begin{align*}
\sum_{k=0}^{n/b} f(kb) B_k
 =  \sum_{i=0}^{n-1} \left ( \sum_{k=\lfloor \frac{i}{b} \rfloor - M' + 1}^{\lfloor \frac{i}{b} \rfloor + M}  f(kb)  c_{i-kb}  \right ) a_{i},
\end{align*}
which can be subtracted from $\sum_{i=0}^{n-1} f(i) a_i$ to prove the result.
\end{proof}

The key idea is that to support fast computations, we want
\begin{align}
\delta (i) = f(i) - \sum_{k=\lfloor \frac{i}{b} \rfloor - M' + 1}^{\lfloor \frac{i}{b} \rfloor + M}  f(kb)  c_{i-kb} = 0
\end{align}
for most integers $i$, as this implies that we can compute almost all
of the range query using only the buffer: i.e., from the previous proposition when $\delta (i) = 0$, we have
\begin{align*}
\sum_{i=0}^{n-1} f(i) a_i & = \sum_{k=0}^{n/b-1} f(kb) B_k.
\end{align*}
It seems remarkable that
the precomputed $B_k$  values are suitable for use with many functions,
possibly including functions not envisioned when the buffer was
initially constructed.

From $\delta (i)=0$ we will arrive at Lagrange interpolation in section~\ref{sec:li},
 since we are mostly interested in the case where $f$ is
locally a polynomial.  Moreover, because we want the ability to store
the buffer component $B_k$ at position $kb$, we also require that
$\delta (kb)=0$. This will ensure that the value $a_{kb}$ is never
needed in Eq.~(\ref{eq:bignasty}).

\begin{proposition} \label{prop:tosatisfyowen} From the definition
\begin{align*}
\delta (i) =
f(i) -\sum_{k=\lfloor \frac{i}{b} \rfloor - M' + 1}^{\lfloor \frac{i}{b} \rfloor + M}  f(kb)  c_{i-kb},
\end{align*}
we have that $\delta (kb)=0$ for all integers $k$ if
\begin{align*}
c_{lb}= \left \lbrace \begin{array}{cl}
 1 & \mathrm{if~} l = 0 \\
 0 & \mathrm{otherwise}\end{array} \right. .
\end{align*}
\end{proposition}
\begin{proof}
Assume that $c_{lb}$ is zero whenever $l\neq 0$, then
\begin{align*}
\delta (lb) & =f(lb) -\sum_{k=l - M' + 1}^{l + M}  f(kb)  c_{(l-k)b} \\
&=f(lb) -  f(lb)  c_{0}
\end{align*}
and the result follows. \end{proof}

Consider the case where $M=0, M'=1$, then
$B_k = \sum_{j=0}^{b-1} c_{j} a_{j+kb}$.
 Suppose we want to buffer
range sums, then $f$ will be 1 except at the endpoints.
So, from $\delta (i) = 0$, we see that we want
\begin{align*}\sum_{k=\lfloor \frac{i}{b} \rfloor }^{\lfloor \frac{i}{b} \rfloor }    c_{i-kb}  = 1\end{align*}
or $c_{i-\lfloor \frac{i}{b} \rfloor b}  = 1$;
that is, $c$ is always 1 within the range of its indices.
 Thus, we retrieve the formula
$B_k = \sum_{j=0}^{b-1} a_{j+kb}$ as the unique solution to buffer
range sums when $M=0, M'=1$. As we shall see, for larger overlaps the solution
is no longer unique and the problem becomes more interesting.
%%%%%%%%%%%%%%%%%%%%%%%%%%%%%%%%%%%%%%%%%%%%%%%%%%%%%%%%%%%%%%%%%%%%%%%%%%%%%%%%%%%%%%%%%%%%%%%%%%
\subsection{An Example: Sum and First Moments ($M=1, M'=1$)}
\label{sec:specialcase}
%%%%%%%%%%%%%%%%%%%%%%%%%%%%%%%%%%%%%%%%%%%%%%%%%%%%%%%%%%%%%%%%%%%%%%%%%%%%%%%%%%%%%%%%%%%%%%%%%%

With $M=0, M'=1$, we can buffer SUM queries.  Using $M=1, M'=1$, we
will buffer the first two local moments: local range sums ($\sum_i
a_i$) and local first moments ($\sum_i i a_i$). %It is important to
 %realize that w
While we will support a wider range of queries, the
buffer size remains unchanged. However, the complexity of the queries does go
up when $N$ increases.

With $M=1, M'=1$, $\delta (i)=0$ implies
\begin{align}
\delta (i) = f(i) - \sum_{k=\lfloor \frac{i}{b} \rfloor }^{\lfloor \frac{i}{b} \rfloor + 1}
 f(kb)  c_{i-kb} = 0.\label{specialcasem1n1}
\end{align}
We recognize this problem as the linear Lagrange interpolation of $f(i)$
using values at $f(\lfloor i / b \rfloor b)$ and $f(\lfloor i / b \rfloor b + b)$.
The next proposition gives a solution to these equations.

\begin{proposition}Given integers $n$ and $b$, such that $b$ divides $n$,
the equation $f(i) - \sum_{k=0}^{n/b-1} f(kb) c_{i-kb} = 0$
 holds for all
linear functions, $f(x)=ax+b$, if $c_{i}=1-\frac{|i|}{b}$
when $i\in [-b,b)$.
\end{proposition}
\begin{proof}
Setting $f(x)=1$ and $f(x)=x$ in Eq.~(\ref{specialcasem1n1})  yields two equations
\begin{equation*}
1 = \sum_{k=\lfloor \frac{i}{b} \rfloor }^{\lfloor \frac{i}{b} \rfloor + 1}   c_{i-kb},~~~~~
i = \sum_{k=\lfloor \frac{i}{b} \rfloor }^{\lfloor \frac{i}{b} \rfloor + 1} kb  c_{i-kb},
\end{equation*}
which are true
when $c_{i}=1-\frac{|i|}{b}$. The general result ($f(x)=ax+b$) follows by linearity.\end{proof}

We can verify that $\delta (kb)=0$, using Proposition~\ref{prop:tosatisfyowen}:
when $l \notin \{0,-1\}$ then $lb \notin [-b,b)$, hence $c_{lb}=0$.  Otherwise,
$c_{lb}=1-\frac{|lb|}{b}$, i.e., 1 when $l=0$ and 0 when $l=-1$.

%%%%%%%%%%%%%%%%%%%%%%%%%%%%%%%%%%%%%%%%%%%%%%%%%%%%%%%%%%%%%%%%%%%%%%%%%%%%%%%%%%%%%%%%%%%%%%%%%%
%%%%%%%%%%%%%%%%%%%%%%%%%%%%%%%%%%%%%%%%%%%%%%%%%%%%%%%%%%%%%%%%%%%%%%%%%%%%%%%%%%%%%%%%%%%%%%%%%%
\section{Overlapped Bin Buffering for Local Moments: \textbf{\textsc{Ola}} Buffers}% and Fundamental Functions
\label{sec:li}
%%%%%%%%%%%%%%%%%%%%%%%%%%%%%%%%%%%%%%%%%%%%%%%%%%%%%%%%%%%%%%%%%%%%%%%%%%%%%%%%%%%%%%%%%%%%%%%%%%
%%%%%%%%%%%%%%%%%%%%%%%%%%%%%%%%%%%%%%%%%%%%%%%%%%%%%%%%%%%%%%%%%%%%%%%%%%%%%%%%%%%%%%%%%%%%%%%%%%

Overlapped Bin Buffering as described in the previous section can be used to buffer local
moments as in subsection~\ref{sec:specialcase}. In the special case of Overlapped Bin Buffering
where the buffers are computed using Lagrange interpolation,
we call the resulting data structure an \textsc{Ola} buffer.

Lagrange interpolation is a common technique
discussed in  standard Numerical~Analysis references~\cite[section 5.5]{rao2002}.
 %Essentially
In essence, given a function $f$ and $M+M'$ samples of the
function $f(m_1),f(m_2),\ldots, f(m_{M+M'})$, then
\begin{enumerate}
\item we solve for the \textbf{unique} polynomial $p$ of degree $M+M'-1$
such that $p(m_1)=f(m_1), p(m_2)=f(m_2),\ldots,p(m_{M+M'})=f(m_{M+M'})$;
\item we evaluate the polynomial $p$ at $x$ and return this as the interpolated value.
\end{enumerate}
It should be evident that if $f$ is itself a polynomial of degree at
most $M+M'-1$, then the interpolation error will be 0;
that is, $p(x)=f(x)$. We say that Lagrange interpolation of order
$M+M'-1$ reproduces polynomials of degree $M+M'-1$.
Also, Lagrange
interpolation is optimal, in the sense that it uses the smallest
possible number of samples while reproducing polynomials of a given
degree.

%In general,
Given $M+M'$ samples $m_1,\ldots, m_{M+M'}$ of a function $f$, the
\emph{Lagrange formula} for
the interpolated value at $x$ is given by
\begin{align}
f(x) = \sum_{k=m_1}^{m_{M+M'}} \left (  \prod_{m=m_1; m \neq k}^{m_{M+M'}}  \frac{x-m}{k-m} \right ) f(k).\label{eq:lagrangeformula}
\end{align}

In Proposition~\ref{propform}, we introduced the function $\delta(i)$ given by
\begin{align*}
f(i) -\sum_{k=\lfloor \frac{i}{b} \rfloor - M' + 1}^{\lfloor \frac{i}{b} \rfloor + M}   c_{i-kb}  f(kb)
\end{align*}
which can also be written as
\begin{align*}
f(i) -\sum_{k= - M' + 1}^{ + M}   c_{r-kb}  f((k+\lfloor \frac{i}{b} \rfloor)b)
\end{align*}
where $r=i-\lfloor \frac{i}{b} \rfloor b$.  On the other hand, as a direct
consequence of the Lagrange formula,
choosing $m_1= b(\lfloor \frac{i}{b} \rfloor -
M' + 1),\ldots$,$m_{M+M'}=b(\lfloor \frac{i}{b} \rfloor + M)$ with $x=i$, we
have
\begin{align*}
f(i) = \sum_{k= - M' + 1}^{ M} \left (  \prod_{m= - M' + 1, m\neq k}^{ M}  \frac{r-mb}{kb-mb} \right ) f((k+\lfloor \frac{i}{b} \rfloor)b).
\end{align*}
Now, by Lagrange's formula, if $f$
is a polynomial of degree $M+M'-1$
and if
\begin{align*}
 c_{r-kb} =\prod_{m= - M' + 1, m\neq k}^{ M}  \frac{r-mb}{kb-mb},
\end{align*}
then $\delta(i)=0.$
When $M=1, M'=1$, Lagrange interpolation becomes equivalent to
linear splines and is trivial to compute: if $k=0$,
$c_{r} =\frac{b- r}{b}$
else if $k=1$,
 $c_{r-b} =  \frac{r}{b}$;
 we conclude that  $c_{r-kb}= 1- \frac{\vert r-kb \vert }{b}$.

For $M+M'>2$, we can simply apply the formula
\begin{align*}
c_{r-kb} &=
\prod_{m= - M' + 1, m\neq k}^{ M}  \frac{r-mb}{kb-mb}\\
&=  \frac{ \prod_{m= - M' + 1, m\neq k}^{ M} r/b-m}{\prod_{m= - M' + 1, m\neq k}^{ M} k-m} \\
&=  \frac{(-1)^{M-k} \prod_{m= - M' + 1, m\neq k}^{ M} r/b-m}{(M-k)!(k+M'-1)!}
\end{align*} and precompute
the coefficients once for $b$ possible values of $r=0,\ldots b-1$ and
$M+M'-1$ possible values of $k=-M'+1,\ldots,M$. This gives a total of $ (M+M'-1) b$ coefficients.

The next lemma applies these results and says that Overlapped Bin
Buffering efficiently buffers the first $M+M'$ moments
when using coefficients derived from Lagrange interpolation.
%Notice that
Hierarchical \textsc{Ola} with $b=\sqrt{n}$ is
One-Scale \textsc{Ola}, thus we need not make an explicit distinction between the two.

\begin{lemma}
\label{lemma:degreeN-queryfcns}
Overlapped Bin Buffering with overlap parameters $M,M' \geq 0$ allows $\delta(i)=0$
for polynomials of degree $M+M'-1$,
if the coefficients $c$ are chosen to be the Lagrange coefficients of degree $M+M'-1$.
\end{lemma}

%%%%%%%%%%%%%%%%%%%%%%%%%%%%%%%%%%%%%%%%%%%%%%%%%%%%%%%%%%%%%%%%%%%%%%%%%%%%%%%%%%%%%%%%%%%%%%%%%%
\subsection{Local Moments Using One-Scale \textsc{Ola}}
\label{section:piecewise}

%%%%%%%%%%%%%%%%%%%%%%%%%%%%%%%%%%%%%%%%%%%%%%%%%%%%%%%%%%%%%%%%%%%%%%%%%%%%%%%%%%%%%%%%%%%%%%%%%%

 We have already seen that
each buffer value $B_k$ in the \textsc{Ola} buffer is %actually
 a sum
over several bins, $B_k = \sum_{j=-Mb}^{M'b-1} c_{j} a_{j+kb}$ where the $c_j$
are determined by Lagrange interpolation or directly as in
subsection~\ref{sec:specialcase}.  By Lemma~\ref{lemma:degreeN-queryfcns},
if $f$ is any polynomial of degree $M+M'-1$,
then local moment queries of the form $\sum_i f(i)a_i$ can be
answered from the \textsc{Ola} buffer alone.
However, consider
function $f(x)=x$ over $[2,10]$ and zero elsewhere. This
function is used when computing the first-order query moment
$\sum_{i=2}^{10} i a_i$.  (See subsection~\ref{section:generalcase}.)
Our approach
must be refined to handle $f$ and other useful functions.

An alternate viewpoint to the approach emphasizes how Lagrange interpolation
provides an approximation $h$ to function $f$.
Knowing how $h$ differs from $f$ allows us to compensate for
$f$'s not being polynomial or (in section~\ref{sec:approxobb})
allows us to bound the
error from imperfect compensation.  The details of this viewpoint follow.

By the proof of proposition~\ref{propform}, if we define
 \begin{align*}h(i) = \sum_{k=\lfloor \frac{i}{b} \rfloor - M' + 1}^{\lfloor \frac{i}{b} \rfloor + M}  f(kb)  c_{i-kb}\end{align*}
then   $\sum h(i)a_i = \sum f(kb)B_k$.
It is interesting to note that given some function $f$,
we can compute $h$ using a Lagrange polynomial \emph{for each bin} as follows.
\begin{enumerate}
\item Pick the bin, and suppose it
comprises
the cells in $[kb,(k+1)b)$.
\item Obtain $M+M'$ data points by sampling $f$ at
$\{(k-M+1)b, \ldots, kb, \ldots, (k+M'-1)b,(k+M')b \}$.
\item There is a unique polynomial $g_{(k)}$ of degree $M+M'-1$ going through all those data points.
 Notice that $g_{(k)}$ is computed bin-wise.
\end{enumerate}
%We observe that w
Whenever $f$ is polynomial of degree at most $M+M'-1$
over $[(k-M+1)b,(k+M')b]$, then $f=g$ over this same interval because the polynomial is unique.
Moreover, the Lagrange coefficients satisfy Proposition~\ref{prop:tosatisfyowen} and
thus $h(kb) = f(kb)$ for all $k$.
Then, we can define a function by piecing together the bin-wise polynomials
and because the Lagrange interpolant is unique, we have
$h_{|[kb,(k+1)b)}=g_{(k)}$, since both sides of the equation are Lagrange interpolants
of the same points.  That is, there is a unique
linear interpolation algorithm of degree $M+M'-1$ over $M+M'$ data points,
exact for polynomials of degree $M+M'-1$.
For our range queries, the query function $f$ is pieced together
from a polynomial (inside the range)
and the polynomial $g(i)=0$ (outside the range).
Hence, $f(i) - h(i)$ will be zero except for two ranges of width $(M+M')b$,
located around each of the range query's end points.
Moreover, $\sum h(i)a_i = \sum f(kb)B_k$ implies by Eq.~(\ref{eq:bignasty}) that
 \begin{align*}
\sum_i \delta (i) a_i &=\sum_i f(i) a_i - \sum_{k=0}^{n/b} f(kb) B_k \\
 &=\sum_i f(i) a_i - \sum h(i)a_i \\
 &= \sum_i (f(i) - h(i)) a_i .
\end{align*}
Therefore $\sum_i \delta (i) a_i$ can be computed in  time $O(N^2b)$,
accessing $M+M'$ bins from our external array.

Hence,  Eq.~(\ref{eq:bignasty}) tells us how to answer the original query,
$\sum_i f(i) a_i$, faster than the $\Omega(n)$ time required without
precomputation. The query can be rewritten as the sum of
$\sum_{k=0}^{n/b} f(kb) B_k$  and $\sum_i \delta (i) a_i$,
(computed in time $O(n/b)$ and  $O(N^2 b)$, respectively).
Therefore,we obtain a net reduction of the complexity from $n$ to $n/b+N^2 b$.

We note that the  $N^2b$ factor could be improved to $Nb$, if we
precompute $h$ for the $b$ possible positions where an endpoint
could fall within a bin, and for the $N$ basic functions
$1, i, i^2, \ldots, i^{N-1}$ we expect used with local moments.
Each of these $Nb$ values would have $\Theta(Nb)$ tabulated entries,
leading to a storage complexity of $\Theta(N^2b^2)$, possibly too high
for the small time savings except for tiny values of $b$.
Yet small values of $b$ would arise in the hierarchical setting that we shall
next explore.

%%%%%%%%%%%%%%%%%%%%%%%%%%%%%%%%%%%%%%%%%%%%%%%%%%%%%%%%%%%%%%%%%%%%%%%%%%%%%%%%%%%%%%%%%%%%%%%%%%
\subsection{Local Moments Using Hierarchical \textsc{Ola}}
\label{section:hierarchicalOLA}
%%%%%%%%%%%%%%%%%%%%%%%%%%%%%%%%%%%%%%%%%%%%%%%%%%%%%%%%%%%%%%%%%%%%%%%%%%%%%%%%%%%%%%%%%%%%%%%%%%

We can further decrease the complexity by a hierarchical approach.
We reduced the complexity from $O(Nn)$ to $O(Nn/b+N^2 b)$.
After the first transform, the cost is dominated by the computation
of $\sum_{k=0}^{n/b} f(kb) B_k$.
By the same method, we can reduce the complexity further to
$n/b^2+N^2 2b$. Because in-place storage is possible, this comes with no
extra storage burden. Applying the method $\log_b n$~times reduces it down
to $O(N^2 b \log_b n)$.
The net result is similar to Bin Buffering, except that we are
able to buffer the first two moments simultaneously, using a single
buffer.

%%%%%%%%%%%%%%%%%%%%%%%%%%%%%%%%%%%%%%%%%%%%%%%%%%%%%%%%%%%%%%%%%%%%%%%%%%%%%%%%%%%%%%%%%%%%%%%%%%
%%%%%%%%%%%%%%%%%%%%%%%%%%%%%%%%%%%%%%%%%%%%%%%%%%%%%%%%%%%%%%%%%%%%%%%%%%%%%%%%%%%%%%%%%%%%%%%%%%
\section{ Using Overlapped Bin Buffering for Fast Local Moments (\textsc{Ola}): Algorithms and Experimental Results }
\label{sec:exp}
%%%%%%%%%%%%%%%%%%%%%%%%%%%%%%%%%%%%%%%%%%%%%%%%%%%%%%%%%%%%%%%%%%%%%%%%%%%%%%%%%%%%%%%%%%%%%%%%%%
%%%%%%%%%%%%%%%%%%%%%%%%%%%%%%%%%%%%%%%%%%%%%%%%%%%%%%%%%%%%%%%%%%%%%%%%%%%%%%%%%%%%%%%%%%%%%%%%%

\structure{Appears to be mostly experiments and results.}

This section puts the ideas of the previous section into practice,
presenting more details of \textsc{Ola} and
presenting an experimental analysis of \textsc{Ola}
as an example of Hierarchical Bin Buffering.
There are three fundamental operations on \textsc{Ola} buffers:
building them initially, using them for fast queries, and finally, updating
them when the underlying data changes.
Each fundamental operation will, in turn, be described and its complexity
analyzed.
However, our complexity analysis
is typical and ignores system-specific factors including the relative costs
of various mathematical operators and the effects of the
memory-access patterns on a computer's memory hierarchy.  To %demonstrate
show that
the algorithm can be efficiently implemented, we
we
coded \textsc{Ola} in C++.  The performance of our
implementation completes the discussion of each fundamental operation.

To enable replication of our results, we next provide some details
of our implementation and the test environment.
Our experiments were conducted with $N$ being even, and
we chose $M=M'=\frac{N}{2}$; these constraints were imposed
by our implementation.  Our test platform was a
Pentium~3 Xeon server with 2~\GiB{} RAM running the Linux operating system
(kernel~2.4), and the software was compiled using the GNU compiler (GCC~3.2
with -O2).
For these experiments, we simulated arrays of any size by
``virtual arrays'' defined by $a_i=\sin(i)$: the function $\sin$
is chosen arbitrarily and the intent is that the
access time to any one array element is a fixed cost (a calculation
rather than a memory or disk access).  Results
are thus less dependent on current disk characteristics than
would be otherwise possible. See section~\ref{sec:external-mem} for disk-based
experiments.

For our experiments, array indices were always 64~bits, whereas stored values
are 32-bit floating-point values.  Unless otherwise specified,
$n \approx 2^{30}$, giving us about 4~\GiB{} of virtual-array data;
by ``size'', we refer to $n$, the number of array elements.
As well, the buffer always fit within main memory
and no paging was observed.  We considered several different
values of $b$: 32, 128, 1024, $2^{15}$ and $2^{20}$.  The first three
values imply hierarchical \textsc{Ola} and can be justified
by the ratio of main memory to external memory on current
machines.    The last two
values imply one-scale \textsc{Ola} and fit the introduction's
scenario that $\sqrt n$ would be an appropriate internal
buffer size ($b=2^{15}$), or fit a scenario where the user
wants whatever gains can be obtained from a tiny buffer.
We chose the value $b=128$ as the ``typical'' value when
one was needed.

For $N$, we experimented with values 2, 4, 8 and 16, with 4 deemed
the typical value.  Value 2 does not enable all the local moments
that are likely to be used in practice, whereas we could not
imagine any scenario where $16^{\mathrm{th}}$ or higher
moments would be useful.

%%%%%%%%%%%%%%%%%%%%%%%%%%%%%%%%%%%%%%%%%%%%%%%%%%%%%%%%%%%%%%%%%%%%%%%%%%%%%%%%%%%%%%%%%%%%%%%%%%
\subsection{Computing the \textsc{Ola} Buffer}
\label{sec:build-algo}
%%%%%%%%%%%%%%%%%%%%%%%%%%%%%%%%%%%%%%%%%%%%%%%%%%%%%%%%%%%%%%%%%%%%%%%%%%%%%%%%%%%%%%%%%%%%%%%%%%

The construction of the \textsc{Ola} Buffer is possible using one pass
over the external array, as illustrated in
Algorithm~\ref{algo:olabufcomp}.

\begin{algorithm}
 {\singlespace
\begin{flushleft}
\textbf{constants:} bin size $b$, even number of buffered moments $N$ and Lagrange coefficients $c$ of degree $N-1$,  $\beta$ is the largest integer such that $n/b^{\beta} \geq N$.
\end{flushleft}
}
 {\singlespace
 \begin{algorithmic}
 \STATE  \textbf{function} computeBuffer($a$):
 \STATE \textbf{INPUT:} an array $a$
 \STATE \textbf{OUTPUT:} an array $B$ (\textsc{Ola} buffer)
 \STATE $B \leftarrow \textrm{onestep}(a,0)$
 \FOR {s = $0,1,\ldots,\beta - 1$ }
 \STATE $B' \leftarrow \textrm{onestep}(B, s)$
 \FOR {$k \in \{0,1,\ldots, size(B')-1\} $}
 \STATE $B_{k b^{s+1}} \leftarrow B'_{k}$
 \ENDFOR
 \ENDFOR
 \end{algorithmic}}
\vspace{5mm}

{\singlespace
 \begin{algorithmic}
 \STATE \textbf{function} onestep($a$,$s$):
 \STATE \textbf{INPUT:} an array $a$
 \STATE \textbf{INPUT:} a scale parameter $s$
 \STATE \textbf{OUTPUT:} an array $B$
 \STATE Allocate $\left \lfloor\frac{size(a)}{b^{s+1}}+1 \right \rfloor$
        components in zero-filled array $B$
 \FOR {$i = 0, 1, \ldots, \lfloor size(a) /b^{s} \rfloor +1$}
  \IF{ $i$ is a multiple of $b$}
  	\STATE $B_{ \lfloor i / b \rfloor } \leftarrow  B_{ \lfloor i / b \rfloor } + a_{i b^{s}}$
   \ELSE
	\FOR{$m = -\frac{N}{2}+1,\ldots, \frac{N}{2}$}
		\STATE $B_{ \lfloor i / b \rfloor +m} \leftarrow B_{ \lfloor i / b \rfloor +m}+ c_{-mb+i \bmod b} a_{i b^{s}}$
	\ENDFOR
   \ENDIF
   \ENDFOR
 \end{algorithmic}}
\caption{\label{algo:olabufcomp}\textsc{Ola} Buffer Computation}

 \end{algorithm}

Each buffer component is over a number of bins that depends
linearly on the number of buffered moments. Similarly, as the size
of the input array increases, we expect a linear increase in the
construction time.  The reason is that
the number of buffered components increases linearly with $n$:
although the number of buffer scales, $\beta$, increases with $n$,
we have $\sum_{k=1}^{\beta} n/b^k \in \Theta(n) $.
Each component has a computation cost that depends only on $N$, leading
to an overall construction time of  $O(Nn)$.

\begin{figure}
\begin{center}
\includegraphics[scale=0.8]{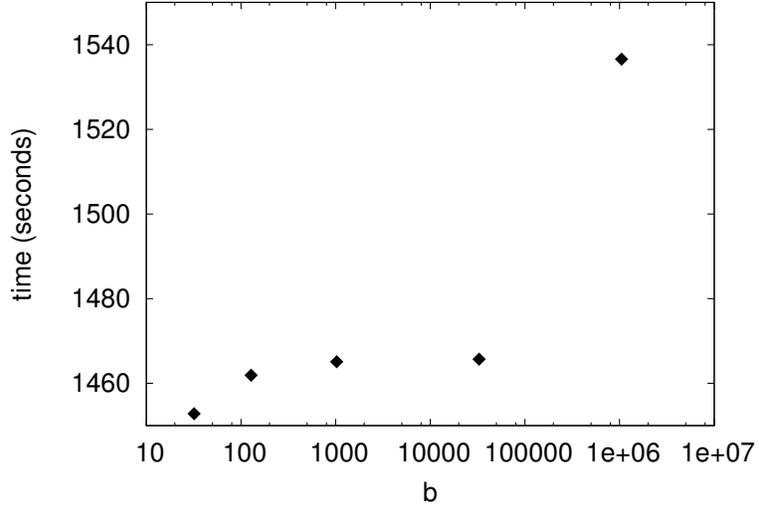}
\end{center}
\caption{\label{fig:construction-b}Time in seconds for the construction
of an \textsc{Ola} buffer, with $N=4$ and varying values of $b$.}
\end{figure}

\begin{figure}[tb]
\begin{center}
\includegraphics[scale=0.8]{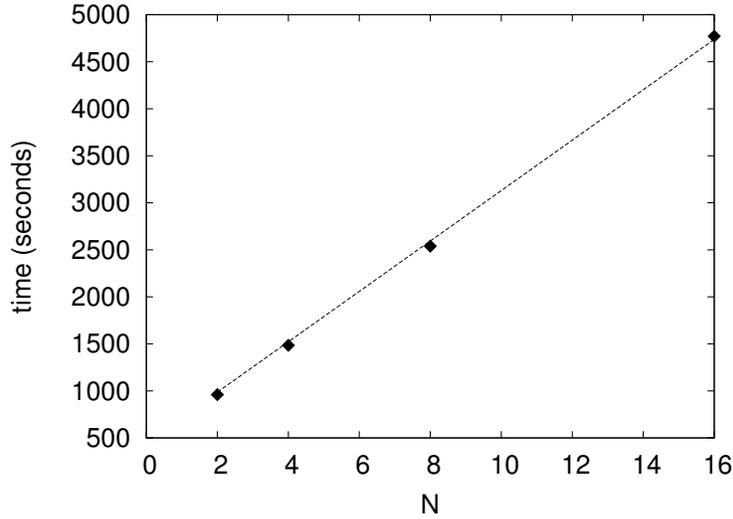}
\end{center}
\caption{\label{fig:construction-N}Time in seconds for the construction
of an \textsc{Ola} buffer, with $b=2^{15}$ (one-scale \textsc{Ola}) and
varying values of $N$. Times were %virtually
almost identical for hierarchical \textsc{Ola}
using $b=128$.  Curve $t(N)=268 N + 450$ (shown) fits the points well. }
\end{figure}

Of course, the storage required is inversely proportional
to $b$: $n/b$. However, we do not expect the construction time
to vary significantly with $b$ as long as the buffer
is internal ($n/b$ is small). Indeed, when $b$ grows, then the
cost of computing each buffer element grows linearly and is
proportional to $Nb$. On the other hand, the number of buffer components
decreases with $1/b$.  In total, the cost is roughly independent of
$b$. If $n/b$ grows substantially, then we might expect
a slight time increase due to poorer memory performance on
the large array. However, for all our experiments, the buffer
size remained much smaller than system RAM\@.
Experimental data to substantiate this is shown in
Figures~\ref{fig:construction-b}--\ref{fig:construction-N}.  %Note that
The
scale of Figure~\ref{fig:construction-b} magnifies what was less than
a 7\% difference between construction times, and %also note that
we saw a
small increase (5\%) in construction time when $b$ increased, which was
not as anticipated.  However, the point stands that buffer
construction was not heavily affected by the choice of $b$.
% On the CoreDuo optiplex, small b=32 is about 3\% faster, and
% b=128 is about 1\% faster, than the larger b values. There is no
% spike at b=1M.  So really nothing to add here.

%%%%%%%%%%%%%%%%%%%%%%%%%%%%%%%%%%%%%%%%%%%%%%%%%%%%%%%%%%%%%%%%%%%%%%%%%%%%%%%%%%%%%%%%%%%%%%%%%%
\subsection{Fast Local Moments Using the \textsc{Ola} Buffer}
\label{sec:query-algo}
%%%%%%%%%%%%%%%%%%%%%%%%%%%%%%%%%%%%%%%%%%%%%%%%%%%%%%%%%%%%%%%%%%%%%%%%%%%%%%%%%%%%%%%%%%%%%%%%%%

The algorithm for fast queries follows from Proposition~\ref{propform}:
recall that we choose $M'=\frac{N}{2}, M=\frac{N}{2}$.
As a first step, we have to compute $\sum_{i} \delta (i) a_i$
where
\begin{equation*}
\delta (i) =
f(i) -\sum_{k=\lfloor \frac{i}{b} \rfloor - \frac{N}{2} + 1}^{\lfloor \frac{i}{b} \rfloor +
\frac{N}{2}}  f(kb)  c_{i-kb}.
\end{equation*}
Then
for scales $s=1,\ldots,\beta$, we add $\sum_{i} \delta^{(s)} (i) B_{i b^{s-1}}$
where
\begin{equation*}
\delta^{(s)} (i) =
f(i b^{s}) -\sum_{k=\lfloor \frac{i}{b} \rfloor - \frac{N}{2} + 1}^{\lfloor \frac{i}{b} \rfloor + \frac{N}{2}}  f(kb^{s+1})  c_{i-kb}.
\end{equation*}
Finally, at scale $s=\beta$,
we add to the previous computations
$\sum_{k=0}^{n/b^{\beta - 1}} f(kb^{\beta+1}) B_{kb^{\beta - 1}}$.

The key to an efficient implementation is to know when $\delta(i)$ ---
or $\delta^{(s)}(i)$ --- will be zero given a function
$f$, so that we only sum over a number of terms proportional to $Nb$
at each scale. In other words, we need to do a lazy
evaluation. Suppose that $f$ is a polynomial of degree at most $N-1$
for $N$ even over the interval $[p,q]$ and the overlap parameters are
$M=\frac{N}{2}, M'=\frac{N}{2}$, then $\delta^{(s)}$ may be nonzero
only over intervals $[p',p'']$ and $[q',q'']$, where $ p' = (\lfloor
\frac{p}{b^{s+1}} \rfloor - \frac{N}{2}) b$, $ p'' = (\lfloor
\frac{p}{b^{s+1}} \rfloor + \frac{N}{2}) b$, $q' = (\lfloor
\frac{q}{b^{s+1}} \rfloor - \frac{N}{2}) b$ and $q'' = (\lfloor
\frac{q}{b^{s+1}} \rfloor + \frac{N}{2}) b$.  The complete pseudocode
is given in Algorithm~\ref{algo:fastmoment}.

\begin{algorithm}[H]

 {\singlespace
\begin{flushleft}
\textbf{constants:} bin size $b$, even number of buffered moments $N$ and Lagrange coefficients $c$ of degree $N-1$,  $\beta$ is the largest integer such that $n/b^{\beta} \geq N$.
\end{flushleft}
}
 {\singlespace
 \begin{algorithmic}
 \STATE  \textbf{function} query($a$, $B$, $f$, $p$,$q$):
 \STATE \textbf{INPUT:} an array $a$
 \STATE \textbf{INPUT:} an \textsc{Ola} buffer $B$
 \STATE \textbf{INPUT:} a function $f$ which is a polynomial of degree at most $N-1$ over $[p,q]$ and zero elsewhere
 \STATE \textbf{OUTPUT:} returns $S=\sum_{i=0}^{n-1} f(i) a_i$
 \STATE $S \leftarrow 0$
 \FOR{$i\in \textrm{candidates}(size(a),0,p,q)$}
 	\STATE $S \leftarrow S + (f(i)-\sum_{k=\lfloor i /b \rfloor - \frac{N}{2} +1}^{\lfloor i /b \rfloor + \frac{N}{2}} f(kb) c_{i-kb}) a_i$
  \ENDFOR
\FOR{$s=1,\ldots, \beta$}
 \FOR{$i\in \textrm{candidates}(size(B)/b^{s-1},s,p,q)$}
 	\STATE $S \leftarrow S + \left (f(ib^s)-\sum_{k=\lfloor \frac{i}{b} \rfloor - \frac{N}{2} +1}^{\lfloor \frac{i}{b} \rfloor + \frac{N}{2}} f(kb^{s+1}) c_{i-kb}\right ) B_{ib^{s-1}}$
  \ENDFOR
\ENDFOR
\STATE $S \leftarrow S + \sum_{k=0}^{size(B)/b^{\beta-1}} f(kb^{\beta+1}) B_{kb^{\beta-1}}$
\end{algorithmic}}
\vspace{5mm}

{\singlespace
 \begin{algorithmic}
 \STATE \textbf{function} candidates($size$,$s$,$p$,$q$):
 \STATE \textbf{OUTPUT:} the union of
 \begin{align*}\left \lbrace \max \left (\left ( \frac{p}{b^{s+1}}-\frac{N}{2}\right)b, 0 \right),\ldots, \left (\frac{p}{b^{s+1}}+\frac{N}{2}\right )b -1 \right \rbrace\end{align*}
 and
\begin{align*}\left \lbrace  \left (\frac{q}{b^{s+1}}-\frac{N}{2}\right )b,\ldots, \min \left (\left (\frac{q}{b^{s+1}}+\frac{N}{2}\right)b-1,size \right ) \right \rbrace.\end{align*}
 \end{algorithmic}}
\caption{\label{algo:fastmoment}Moment Computation using the \textsc{Ola} Buffer}
 \end{algorithm}

Queries are $O(b \beta)$ where $\beta = \log_b n$,
so when $b$ increases the algorithm's running
time increases in proportion to $\frac{b}{\log b}$.
Therefore,
because the buffer size is given by $n/b$, the algorithm becomes slower
as the size of the buffer is reduced.
Experimentally, this was measured by randomly selecting, with
replacement, 2000 of the
$n \choose 2$
different non-empty
ranges. More precisely, we choose two uniformly distributed random numbers $a$ and $b$
and pick the interval $[\min (a,b), \max (a,b))$. We set $N$ to a ``typical value'' of 4, and timed the 2000 sums
and 2000 first moments for various $b$ values\footnote
{ When $N^2 b$ was large, we tested only
$\lfloor \frac{800,000}{N^2b} \rfloor$ cases, to keep test times reasonable.}.

\begin{figure}
\begin{center}
\includegraphics[scale=0.8]{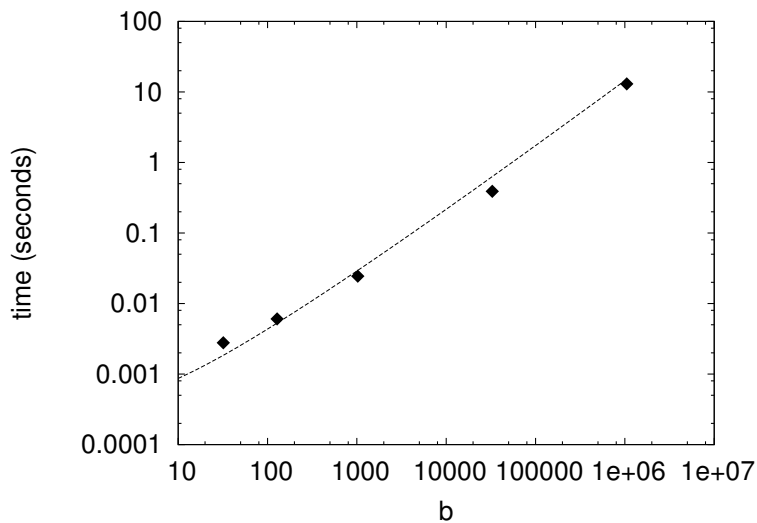}
\end{center}
\caption{\label{fig:time-varying-b} Average time (in seconds)
for the computation
of many randomly selected range sums for $N=4$ and various values of $b$.
Function $t(b)=b/(5000 \ln b)$ is also shown.
}
\end{figure}

Results are shown
in Figure~\ref{fig:time-varying-b};  we also plotted the function
$t(b)=b/(5000 \ln b)$.  The measured running time appears to
grow no faster than $t(b)$.

\begin{figure}[H]
\begin{center}
\includegraphics[scale=0.8]{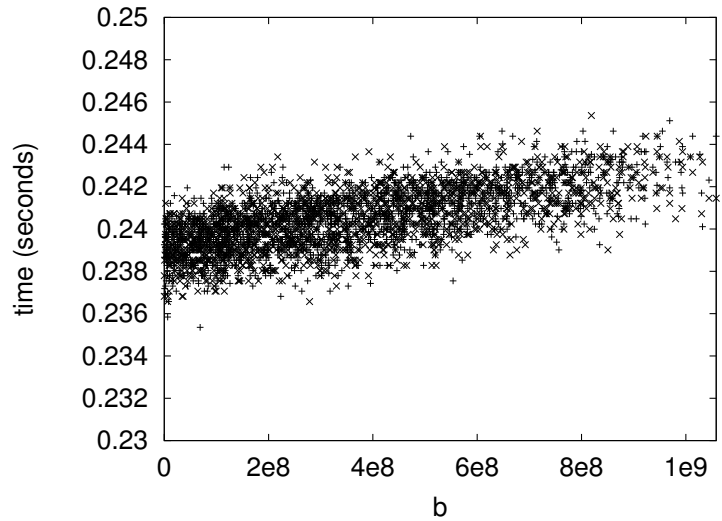}
\end{center}
\caption{\label{fig:time-varying-range32k}Average time (in seconds)
 versus range length for the computation
of 1,525 randomly selected first moments  (x) and 1,525 sums (+)
for $N=4$ and $b=2^{15}$. (One-Scale \textsc{Ola}.)
}
\end{figure}

Note that the time for a range query is affected somewhat by the
length of the range $l$, in that the number of buffer elements $B_i$
accessed will be approximately $l/b$ for one-scale \textsc{Ola}.
(For hierarchical \textsc{Ola} the relationship between the number of
buffers accessed and the range length is much more complex.)
As well, for hierarchical
\textsc{Ola}, the number of hierarchical levels processed will also
depend on the precise positioning of the range's endpoints.  To see
these effects, we plotted the time\footnote{Timed on a slightly faster
Pentium~4 machine with 512~\MiB{} RAM, running a Linux 2.4 kernel that had
been patched to supply high-resolution timings and
hardware performance counts via
PAPI\cite{brow:papi-hpca}.} versus the range size.  Results for
One-Scale \textsc{Ola} (see Figure~\ref{fig:time-varying-range32k})
are as expected: the time was dominated by the (unvarying) work done
around the range's endpoints.  There was a small additional
contribution coming from the number of buffer values accessed, which
showed up as a slight upward slope on the cluster.  First moments and
sums behaved similarly.

\begin{figure}[H]
\begin{center}
\includegraphics[scale=0.8]{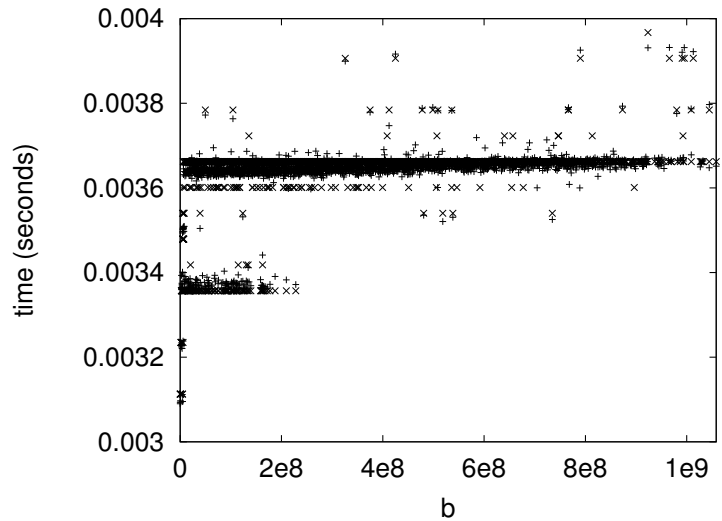}
\end{center}
\caption{\label{fig:time-varying-range128}Time in seconds versus range length for
 2,000 randomly selected first moments (x) and 2,000 sums (+)
with $N=4$ and $b=128$ (4-Scale \textsc{Ola}).}
\end{figure}

The situation is more complex for hierarchical \textsc{Ola} (see
Figure~\ref{fig:time-varying-range128}), where the positioning of the
range determines the number of buffer values and where the positioning
of each endpoint determines how many external array elements are
accessed and determines
how many hierarchical scales need to be considered for the region
surrounding each endpoint.
Since the running times are smaller, it is
perhaps not surprising that the data appears noisier.

\begin{figure}
\begin{center}
\includegraphics[scale=0.8]{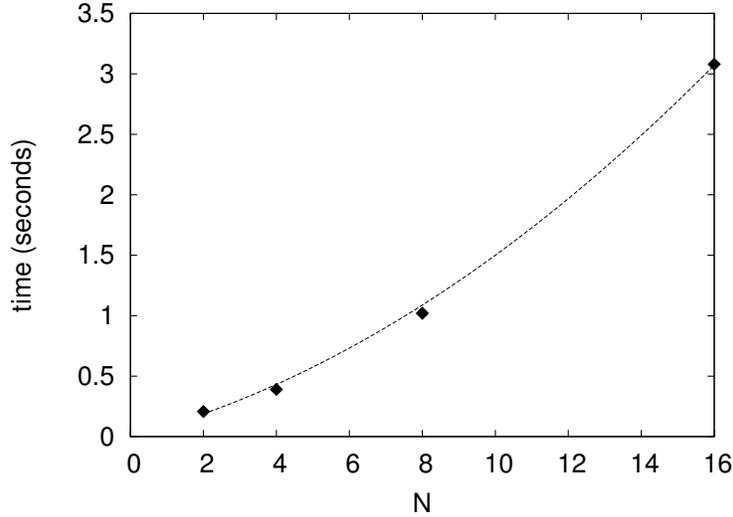}
\end{center}
\caption{\label{fig:querytime-vs-N-onelevel}
Average time per query (seconds) versus $N$ for randomly selected range sums
with $b=2^{15}$ (one-scale \textsc{Ola}).
(The points fit $t(N) = .007N^2+.08N$ well.)}
\end{figure}

\begin{figure}
\begin{center}
\includegraphics[scale=0.8]{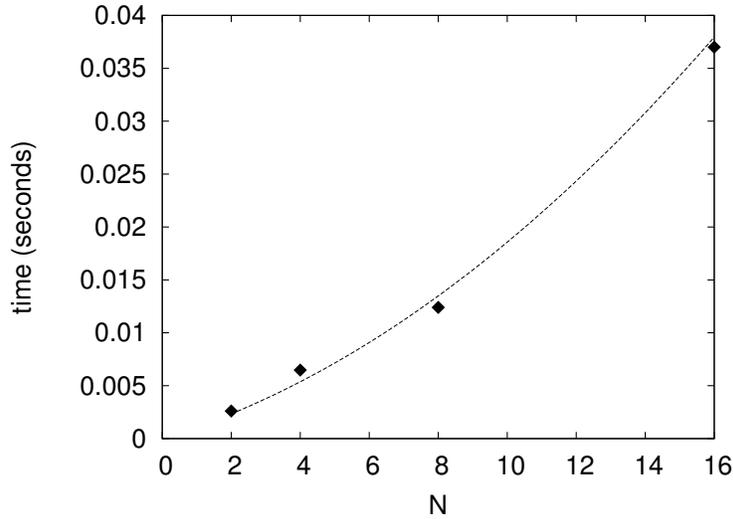}
\end{center}
\caption{\label{fig:querytime-vs-N-multilevel}
Average time per query versus $N$ for 2000 randomly selected range sums
with $b=32$ (3- or 4-Scale \textsc{Ola}).
The running time fits $t(N) = (N^2 + 11.7N)/11700$ well.
}
\end{figure}

We can show that query times for hierarchical \textsc{Ola} grow
quadratically as the number of moments buffered is increased.
For one-scale \textsc{Ola}, queries have two main sources of cost:
first, the cost from computing $\sum_k f(k) B_k$, where $f$ is
piecewise 0 or an $N-1^{\mathrm{st}}$
degree polynomial, which we evaluate at a cost of $\Theta(N)$ per
point within the range of the query.  The expected range of
our queries is long, so this cost is significant.  The second
cost of our queries comes from a $\Theta(N^2)$ calculation done
around the range's endpoints.  Therefore, for
our small values of $N$, the total cost includes
both a large $N^2$ as well as a large linear component.  From
a theoretical point of view, however, the growth is $\Theta(N^2)$ and
is dominated by the endpoint computations.
(See Figures~\ref{fig:querytime-vs-N-onelevel}~and~\ref{fig:querytime-vs-N-multilevel}).
Hence, it might be detrimental to buffer many more moments than we require.
However, the number of moments has no effect on the space complexity,
unlike the bin size, $b$.  Therefore, for large enough arrays, even if
we buffer many moments, the \textsc{Ola} approach will still be
several order of magnitude faster than unbuffered queries.

The \textsc{Ola} approach was not %generally
sensitive to the query:
range sums or first moments were measured to take
almost exactly (within 1\%) the same time.  Therefore,
these results are not plotted.

Based on the theoretical analysis, %it is clear that
\textsc{Ola} can
permit huge query speedups, given extreme values for parameters such
as the relative speeds of internal versus external memory,
amount of memory allocated to the buffer, and so forth.  However, we
need good speedups for ``reasonable'' parameters.  From our
experiments, it is evident that buffered arrays were considerably
faster, and random queries that averaged about
71.6~s without buffering
could be answered in 0.390~s or
0.00605~s when the 4~\GiB{} dataset was
 %respectively
buffered with 128~\KiB{} or 32~\MiB{} (using N=4).  This
corresponds to respective speedups of 184 and 11800.  The construction
time of approximately 1500~s
means a total construction+query
break-even is achieved after about \owensilent{1500/72}
21 queries.

%%%%%%%%%%%%%%%%%%%%%%%%%%%%%%%%%%%%%%%%%%%%%%%%%%%%%%%%%%%%%%%%%%%%%%%%%%%%%%%%%%%%%%%%%%%%%%%%%%
\subsection{Updating the \textsc{Ola} Buffer}
\label{sec:update-algo}
%%%%%%%%%%%%%%%%%%%%%%%%%%%%%%%%%%%%%%%%%%%%%%%%%%%%%%%%%%%%%%%%%%%%%%%%%%%%%%%%%%%%%%%%%%%%%%%%%%

To update the buffer, we can consider how the buffer was originally
constructed: it was computed from the data source and then, in a
hierarchical manner, buffers were computed from the previous buffer
and stored in place.  Recall, for instance, that in the \textsc{Ola}
buffer, values of the second-scale buffer are stored in indices $b,
2b, \ldots, (b-1)b, (b+1)b, (b+2)b, \ldots, (2b-1)b, (2b+1)b, \ldots$
whereas the values of the third-scale buffer are stored at $b^2, 2b^2,
\ldots, (b-1)b^2, (b+1)b^2, \ldots$ and so forth.  We define the cells
at scale $s$ as those having an index divisible by $b^{s-1}$, with the
expository convention that ``scale 0'' refers to entries in the
external array. Our updates propagate changes from smaller scales to
larger scales.  (See Algorithm~\ref{algo:updatenonrecursive}.)

\begin{figure}[H]
\begin{center}
\includegraphics[width=0.95\textwidth]{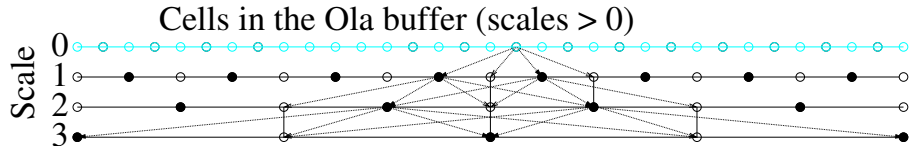}
\end{center}
\caption{\label{fig:updatetree}
Update dag for \textsc{Ola} buffer with $M=M'=2$ ($N=M+M'=4$)
and $b=2$ (see Algorithm~\ref{algo:updatenonrecursive}). Each column
with an entry at Scale 1 corresponds to a buffer cell, whereas
entries at Scale 0 are in the external array.
}
\end{figure}

To understand the update algorithm acting upon this hierarchical
buffer with in-place storage, it is helpful to consider the
``is computed from'' relation between cells, which forms a
directed acyclic graph (dag).  Showing each cell at every scale
to which it belongs, coloring (black) the largest scale for each
cell,  and focusing only on the
part of the dag that needs to be updated, we obtain
Figure~\ref{fig:updatetree}.
The black vertices at scale $s$ (for $s \geq 1$) in the dag correspond to indices
$i$ such that $\lfloor \frac i {b^{s-1}} \rfloor $ is not divisible by $b$, that
is, cells that do not belong to scale $s+1$. The uncolored vertices belong
to scale $s+1$ and in-place storage
means that an update %actually
affects the black node beneath it (except for the
largest scale since the algorithm terminates).
The portion of the dag that is reachable from the changed cell (at
scale 0) is called the \emph{update dag}.

From this, we see that the update cost is
linear with the height of the update dag.  %In order to
To prove it, we first
observe that we can bound, independently of the height of the dag
(given by $\beta \sim \log_b n$), the number of cells per scale that need to be updated.

\begin{proposition}By Algorithm~\ref{algo:updatenonrecursive},
given an update of one cell in the external array, updates are
propagated from one scale to another over at most
$2N$ cells.  In other words, the update dag (as in
Figure~\ref{fig:updatetree}) has at most $2N$ nodes at each level.
\end{proposition}

\begin{proof}
Let $l_{(s)}$ be the difference in indices between the last modified
buffer cell and the first modified one (ordering is by indices) at the end of
step $s$ in Algorithm~\ref{algo:updatenonrecursive}.
In other words, $l_{(s)}$ is the ``range'' of the modified
cells at step $s$.  By convention, $l_{(0)}=0$.  From the algorithm, we
see that $l_{(s)} \leq l_{(s-1)} + N b^{s}$. Hence, we have that
$l_{(s)} \leq 2 N b^{s}$ so that $l_{(s)}/ b^{s} = 2 N$. Hence, each time
we move from one scale to another, at most $2N$ cell values are modified.
\end{proof}

This proposition tells us that the middle  \texttt{for} loop  in
Algorithm~\ref{algo:updatenonrecursive} has at most $2N$ steps; since we do $O(N)$
operations within each, the update complexity is $O(N^2 \beta)$.

Correctness of the algorithm is straightforward and relies on the
fact that index $i$ is processed only at its largest scale (see
the first \texttt{if} statement).  At this time, all updates to
$i$ will have been completed.

\begin{figure}[H]
\begin{center}
\includegraphics[scale=0.8]{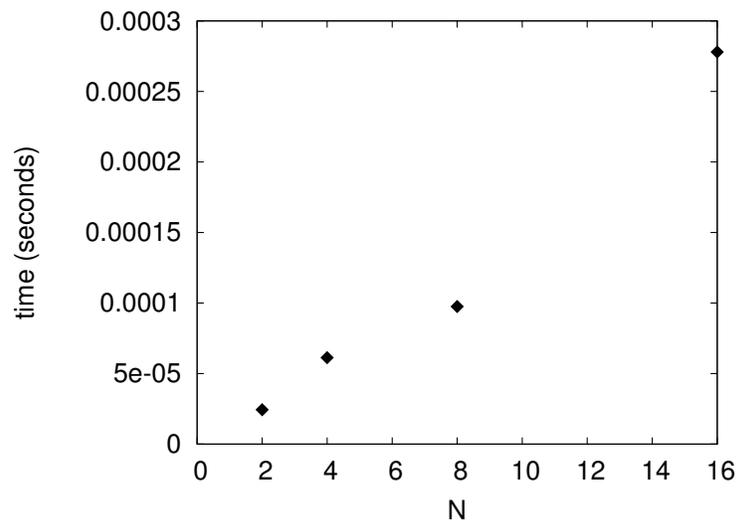}
\end{center}
\caption{\label{fig:update-vs-N-multilevel}
Average time versus $N$ for 200,000 random updates
with $b=128$ (3- or 4-Scale \textsc{Ola}).
}
\end{figure}

The effect of $N$ on computational cost as measured experimentally is given by
Figure~\ref{fig:update-vs-N-multilevel}.  The observed relationship appears
linear, but the three collinear points are misleading; for $N=2$ and $N=4$,
we had $\beta = 4$.  However, for $N=8$ and $N=16$, we had $\beta=3$.
As well, since only $O(N)$ hashmap entries are created (and then updated
$O(N)$ times each), if hashmap-entry creation is expensive, then
the running times will contain a large linear component in $N$.

If the original array was not dense, that is, if most components were
zero, then it can be more efficient to construct the buffer
starting with a zero buffer and then adding each non-zero value
as an update. Because the cost of each update is $O(N^2 \beta)$, if there
are $d(n)$ non-zero values in the original array, then the complexity
of building the buffer through updates is $O(d(n) N^2 \beta)$.
This is asymptotically better than Algorithm~\ref{algo:olabufcomp} whenever
$d(n) \in o(n / N \beta)$.  Experimentally, for $N=4$ and $b=128$,
we made 200k random updates in about 12 seconds.  Thus, even if data items
in a sparse set were added in an unordered manner, it would be faster
to build the buffer through updates if it had about 24~million or fewer
elements,  %200k x (1462/12)
or a density of $\frac{2.4 \times 10^{7}}{1 \times 10^9}$ = 2.4\% or less.
Since update time decreases rapidly with $b$
whereas the time for Algorithm~\ref{algo:olabufcomp}
is almost independent of $b$,  once $b\geq 2^{15}$ incremental
construction is a reasonable alternative to Algorithm~\ref{algo:olabufcomp}
for any data set.

\begin{algorithm}[H]

 {\singlespace
\begin{flushleft}
\textbf{constants:} bin size $b$, even number of buffered moments $N$ and
Lagrange coefficients $c$ of degree $N-1$,
$\beta$ is the largest integer such that $n/b^{\beta} \geq N$.
\end{flushleft}
}

{\singlespace
 \begin{algorithmic}
 \STATE \textbf{function} update ($B$, $j$, $\Delta$):
 \STATE \textbf{INPUT:} an index $j$ in the original array $a$
 \STATE \textbf{INPUT:} an \textsc{Ola} buffer $B$ over the array $a$
 \STATE \textbf{INPUT:} the change $\Delta$ in the value of $a_{j}$
 \STATE \textbf{RETURN:} modifies $B$

  \STATE $deltas$ is a (hash) map \COMMENT{assume 0 for unassigned values}
  \STATE  $deltas_j  \leftarrow \Delta$
  \FOR {$s = 0, \ldots, \beta $}
    \STATE Let $\mathrm{keys}(deltas)$ be the set of keys for the hash table $deltas$ at this point
    \STATE \COMMENT{ Invariant: $\mathrm{keys}(delta)$ contains only indices at scale $s$}
  \FOR {$i \in \mathrm{keys}(deltas)$}
    \IF {$\lfloor \frac i {b^s} \rfloor $ is not divisible by $b$ }

         \STATE \COMMENT{Process $i$ because $s$ is its largest scale}
	 \STATE $\delta \leftarrow deltas_i$
         \FOR{$m = -\frac{N}{2}+1,\ldots, \frac{N}{2}$}
                \STATE $deltas_{(\lfloor \frac{i}{b^{s+1}} \rfloor+m) b^{s+1}} \leftarrow deltas_{(\lfloor \frac{i}{b^{s+1}} \rfloor+m) b^{s+1}} + c_{-bm+ (\lfloor \frac{i}{b^s} \rfloor \bmod b) } \delta$
         \ENDFOR
	 \IF{$i$ is divisible by $b$}
	 \STATE  \COMMENT{ Only (possibly) $j$ is not a multiple of $b$}
         \STATE $B_{i/b}\leftarrow B_{i/b} + deltas_{i}$
         \ENDIF
	 \STATE remove key $i$ from $deltas$
  \ENDIF
  \ENDFOR
  \ENDFOR
  \STATE
  \COMMENT{ Cells belonging to scales $\beta+1$ and above are still in $delta$
            and need to be added to the buffer (see uncolored nodes at the last level of Figure~\ref{fig:updatetree}).}
  \FOR {$i \in \mathrm{keys}(deltas)$}
	 \STATE $B_{i/b}\leftarrow B_{i/b} + deltas_{i}$
  \ENDFOR
  \end{algorithmic}}
 \caption{\label{algo:updatenonrecursive}Updating the \textsc{Ola} Buffer.}
 \end{algorithm}

\begin{figure}[H]
\begin{center}
\includegraphics[scale=0.8]{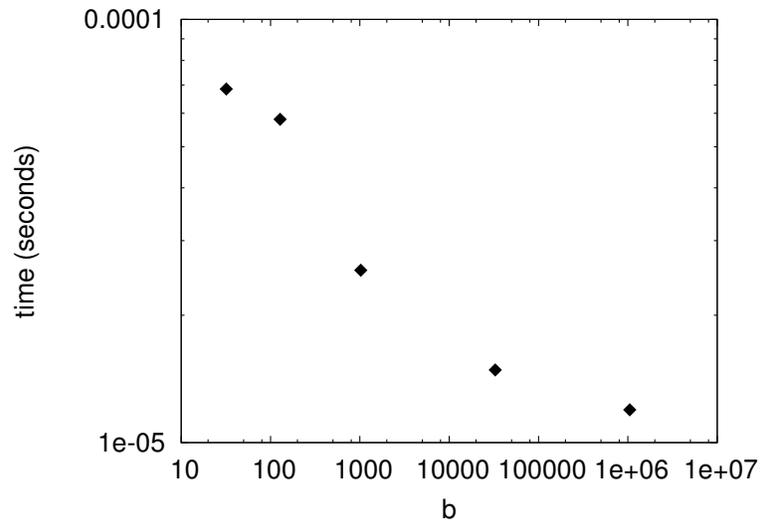}
\end{center}
\caption{\label{fig:update-vs-b}
Average time versus $b$ for 200,000 random updates
with $N=4$
}
\end{figure}

A key point is that updates to the buffer get progressively less
expensive as $b$ goes up and the size of the buffer goes down.
Figures~\ref{fig:update-vs-b} and \ref{fig:update-vs-N-multilevel}, as well
as Table~\ref{tab:update-vs-beta}  provide
experimental evidence of these claims.

\ifthenelse{\boolean{acmformat}}{
\begin{acmtable}{0.335\columnwidth}
\centering
\begin{tabular}{|l|l|l|l|}
$b$      & $\beta$ & time ($\mu s$) & time/$\beta$ \\ \hline
32       & 5 & 68.5                 &  13.7 \\
128      & 4 & 58.1                 &  14.5 \\
1024     & 2 & 25.6                 &  12.8 \\
$2^{15}$ & 1 & 14.9                 &  14.9 \\
$2^{20}$ & 1 & 12.0                 &  12.0 \\
\end{tabular}
\caption{\label{tab:update-vs-beta}
Linear relationship observed between $\beta$ and update time.  $N=4$ and
average time was over 200,000 random updates.}
\end{acmtable}
}{
\begin{table}
\caption{\label{tab:update-vs-beta}
Linear relationship observed between $\beta$ and update time.  $N=4$ and
average time was over 200,000 random updates.}
\begin{center}
\begin{tabular}{|l|l|l|l|}\hline
$b$      & $\beta$ & time ($\mu s$) & time/$\beta$ \\ \hline
32       & 5 & 68.5                 &  13.7 \\
128      & 4 & 58.1                 &  14.5 \\
1024     & 2 & 25.6                 &  12.8 \\
$2^{15}$ & 1 & 14.9                 &  14.9 \\
$2^{20}$ & 1 & 12.0                 &  12.0 \\ \hline
\end{tabular}
\end{center}
\end{table}
}

%%%%%%%%%%%%%%%%%%%%%%%%%%%%%%%%%%%%%%%%%%%%%%%%%%%%%%%%%%%%%%%%%%%%%%%%%%%%%%%%%%%%%%%%%%%%%%%%%%
\subsection{External Memory}
\label{sec:external-mem}

The use of ``virtual arrays'' in the previous section allowed us to abstract
away from the specific details of current memory-system hardware.
However, one might %we can %it is possible to
 abuse such simplified models,
% and incorrectly predict
thus incorrectly predicting
% owen 1. takes the guilt of abuse from us ; 2. tries to get rid of
% the comma-separated prediction because we shouldn't use the
% comma if it shares the subject with the first part....but the
% sentence is complex and otherwise wants some punctuation
good practical performance for algorithms that make
irregular and non-local accesses to disk.  Nevertheless, our algorithms
for buffer construction and queries tend to have good locality.
For instance, with queries in One-Scale \textsc{Ola}, two consecutive groups of
indices (around either endpoint of the queried range) are accessed.
Experiments to support our claims were derived using memory-mapped
files.  Unfortunately, due to our experimental setup (mainly a 32-bit address
space), we were forced to choose a smaller value of
$n \approx 2^{28}$ elements, or about
1~\GiB{} of data.  These experiments were performed
on a computer with 512~\MiB{} of RAM, and since much I/O was
anticipated (and observed), we took wall-clock times while the system
ran in single-user mode.

\begin{figure}[H]
\begin{center}
\includegraphics[scale=0.8]{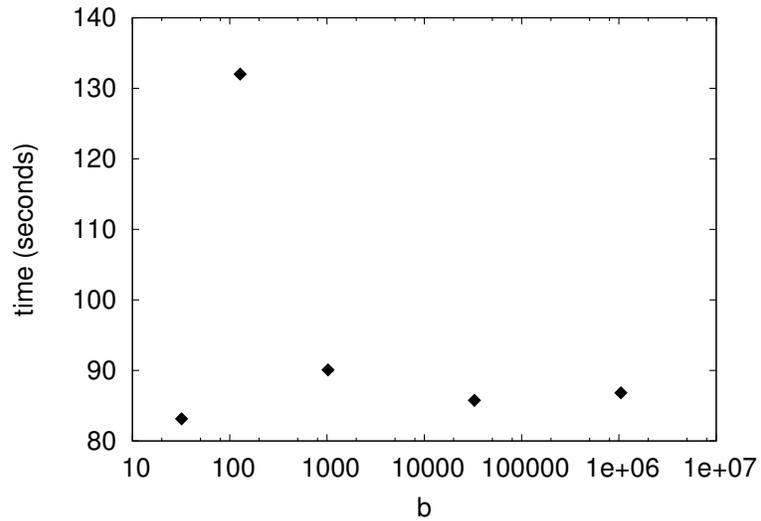}
\end{center}
\caption{\label{fig:extern-constr-vs-b}
Time to construct an \textsc{Ola} buffer ($N=4$) from a memory-mapped disk
file, versus $b$.
}
\end{figure}

Repeating the experiments in which we timed the construction of an \textsc{Ola}
buffer with $N=4$ and varying $b$, we obtained the results shown
in Figure~\ref{fig:extern-constr-vs-b}.  We note that
the discrepancy for $b=128$ does not seem to be an error: it
was repeatable.   Except for this one value of $b$, we see that changes in $b$ affected
construction time by less than 10\%.
The large discrepancy at $b=128$ apparently came from
the operating system and system libraries\footnote
{Using the same hardware, but with the Linux kernel upgraded to version
% I am not sure I know what the kernel, glibc and C++ versions were in 2004
% the text says we were using a 2.4.x kernel, though...
2.6.20, glibc to version 2.5, and the GNU C++ compiler to version 4.1.2,
we obtained different results: the cases $b=32$, $b=128$ and $b=2^{20}$
were similar.
(The median time of 25 runs for $b=128$ was no more
than 5\% larger than the medians of the other two cases.)  The cases of
$b=1024$ and $b=32768$ were similar, their medians being slightly less than
20\%  faster than $b=128$.  Repeating tests for $N=16$,
all cases except $b=2^{20}$ were similar to one another, whereas $b=2^{20}$ was
approximately 50\% slower than the others.  The effects of varying $N$
and $b$ are described in Section~\ref{sec:build-algo}; we conjecture
that some combinations of $N$ and $b$ produce page-access streams
that are easier for the operating system to handle efficiently.
Further investigation is outside the scope of this paper.
}.

By conducting the experiments of subsections
\ref{sec:build-algo}~to~\ref{sec:update-algo} with virtual arrays, we
avoided many secondary system-level effects that might have
obscured our results.  But it is useful to compare the timings
obtained with (realistic) memory-mapped array versus those from our
synthetic virtual arrays. For reference, with virtual arrays
and $n \approx 2^{28}$, a construction time of about 250~s
 was obtained with $N=4$,
for all values of $b$.
We see that the virtual array lead to a construction time
that was approximately three times longer.

\begin{figure}[H]
\begin{center}
\includegraphics[scale=0.8]{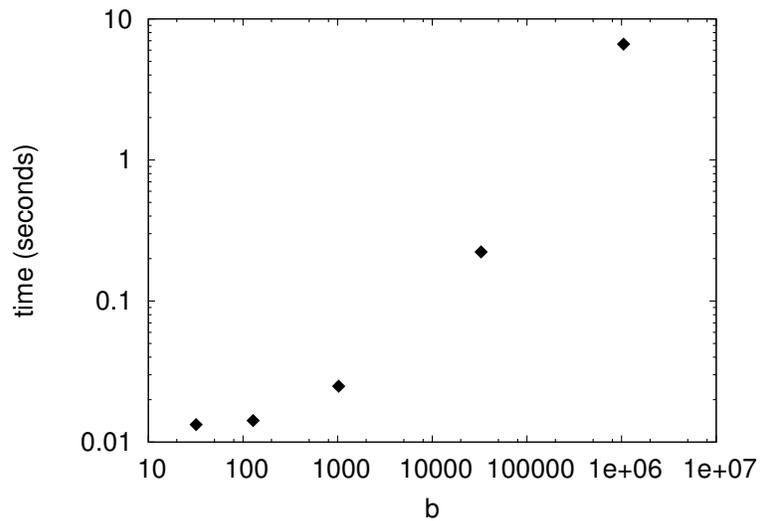}
\end{center}
\caption{\label{fig:extern-query-vs-b}
Average time to answer a random range-sum query from an \textsc{Ola} buffer ($N=4$) versus $b$.
A memory-mapped file with $n \approx 2^{28}$ 4-byte floating-point numbers was used.
}

\end{figure}

We also timed random range-sum queries using our memory-mapped array
(see Figure~\ref{fig:extern-query-vs-b}).  For comparison, similar
random queries were also computed directly from the external array, without
using the \textsc{Ola} buffer at all.  Despite the good locality
of the obvious algorithm for this task, with $N=4$ and $b=128$,
\textsc{Ola} answered the query less than .014 seconds, versus 6.3 seconds
when no buffer was used: a speedup of over 400.
For reference, a virtual array lead to query times that were
approximately 11\%
\owensilent{1525 queries in 376~s sine vs 339.5~s mmap  with -O2}
\emph{slower} than with the memory-mapped implementation for $b=2^{15},\  N=4$
 but about 75\%  faster for
$b=128,\  N=4$. \owensilent{2000 queries in 7.4~s virtual, 28.4~s mmap with -O2}

Comparing Figure~\ref{fig:extern-query-vs-b} to Figure~\ref{fig:time-varying-b},
we observe that with the memory-mapped array, the query
time is not as sensitive to differences in $b$, when $b$
is small. Presumably this is
due to blocking on disks: even when $b$ is small, at least an
entire virtual memory page or disk block needs to be
dedicated to the area around each endpoint of the
query range.

%%%%%%%%%%%%%%%%%%%%%%%%%%%%%%%%%%%%%%%%%%%%%%%%%%%%%%%%%%%%%%%%%%%%%%%%%%%%%%%%%%%%%%%%%%%%%%%%%%

\section{\textsc{Ola} versus \textsc{Bin Buffering}}
\label{sec:obbvshbb}

Assume that we are given the task of buffering
$N$~moments with a fixed amount of internal memory $K$
over a very large array of size $n$. Recall that given an
array of size $n$ and a buffer size $b$, \textsc{Ola} will use
a buffer of size $n/b+1$. Hence,
\textsc{Ola} would lead to  bins of size $b=n/(K-1) \approx n/K$
whereas \textsc{Bin Buffering} would use larger bins of size
$b'=N n /K$.

Assume we can read bins of size $b$ from
external memory with a fixed cost of $E_b$~units of time, and we can access
internal memory cells with a cost of 1~unit of time.
For simplicity, we also assume that $Nb < E_b$, which seems
likely given the small values of $N$ anticipated.
One-scale \textsc{Ola} and \textsc{Bin Buffering} have then
exactly the same complexity, that is, queries have
worst-case complexity
$O(N E_b + K)$. The hierarchical versions also have similar
complexity to one another.

However, not all queries have the same cost: the two algorithms
are \textbf{not} equivalent. \textsc{Bin Buffering}
will support ${K/N} \choose 2$
ranges without \emph{any} access to the external array: all range queries from
bin edges to bin edges can be answered entirely from the buffer.
For instance, $\sum_{i=2K/N}^{10K/N-1}a_i$ can be answered by
summing 8 buffer elements.

However, for some applications, we might be interested in how well
we can approximate the query without access to the external array.
This is especially important in applications such as visualization,
where a very fast initial approximation is valuable.

To explain why \textsc{Ola} is more competitive in providing good
approximations using only the buffer, take the case where $N=2$
and assume that the values in the external array are uniformly bounded
in absolute value by $\kappa$. That is, $\vert a_j \vert \leq \kappa$ for all $j$.
Recall that we assume that the internal buffer has a size of $K$.
Then consider range sums such as $\sum_{j=k}^{l} a_j$. The largest
error made by \textsc{Bin Buffering} is bounded by $2 b' \kappa  =4 \kappa  n /K$ since we
miss at most one bin at each end (whose total value is at most $b' \kappa$).
We can %actually
reduce this bound
to $2 \kappa n / K$ by choosing
to add bins whenever they are more than half occupied.
On the other hand, the largest error that \textsc{Ola} can make is
bounded by $2 \kappa (b/2) =  \kappa n /K$.
Indeed, the worst error is reached when range edges match with bin edges.
In that case, we wrongly take a full bin at each end.
Actually, due to the linear decrease in $c_i$ values, the more distant
values in these bins are weighted lightly; this leads to
the $b/2 \kappa$ bound on each bin.
Hence, \textsc{Ola} is twice as accurate for estimating range sums from the buffer when $N=2$.
Irrespective of $N$, the error for \textsc{Ola} is more than bounded by
$2 b \kappa = 2 \kappa n /K$.
However, the error made by \textsc{Bin Buffering}
can be as bad as $ b' \kappa  = N \kappa  n /K$, even taking into account the possible
improvement one gets by including bins that are more than half used. In other words, as $N$ grows,
the worst-case error made by \textsc{Bin Buffering} grows linearly, unlike
\textsc{Ola}. This result applies to hierarchical versions of these algorithms as well.

Hence, one might want to look at the case $N=8$. This is not unreasonable in
a visualization setting where the user can set the degree of the polynomials to be
fitted. In such a case, \textsc{Bin Buffering} has very large bins
(8 times larger than \textsc{Ola}) which might be undesirable: the approximation
power of \textsc{Bin Buffering} for range sums is at least 4 times lower than
\textsc{Ola} because of the much larger bins.

\section{Approximate Queries using \textsc{Ola}}
\label{sec:approxobb}
\structure{put wiggleresult.tex inline}

We have %previously
seen that \textsc{Ola} can have a
competitive advantage when we are interested in getting approximate
queries out of the memory buffer.
Indeed, \textsc{Ola} supports a wide
range of query types using a single memory buffer and relatively small
bins. However, as with wavelet-based
techniques~\cite{Vitterdatacubeapprox,Approximatewavelets,propolyne},
\textsc{Ola} can support incrementally
better estimates. With wavelet-based methods, one gets approximations
by selecting the most significant wavelet coefficients.  However,
this approach calls for storing all coefficients in order %owen reverted for flow
to
quickly answer queries within a user-specified error bound.
This would be unacceptable for many applications: the
wavelet buffer is as large as the external array itself. Incrementally
better \textsc{Ola} approximations can be computed by first using the
internal buffer, and then adding bins one by one. Indeed, for a given
range query, the \textsc{Ola} algorithms involve many bins at both
ends of the range. However,  only the first few
bins have a significant contribution.

\begin{figure}
 \begin{center}
\includegraphics[angle=270,width=10cm]{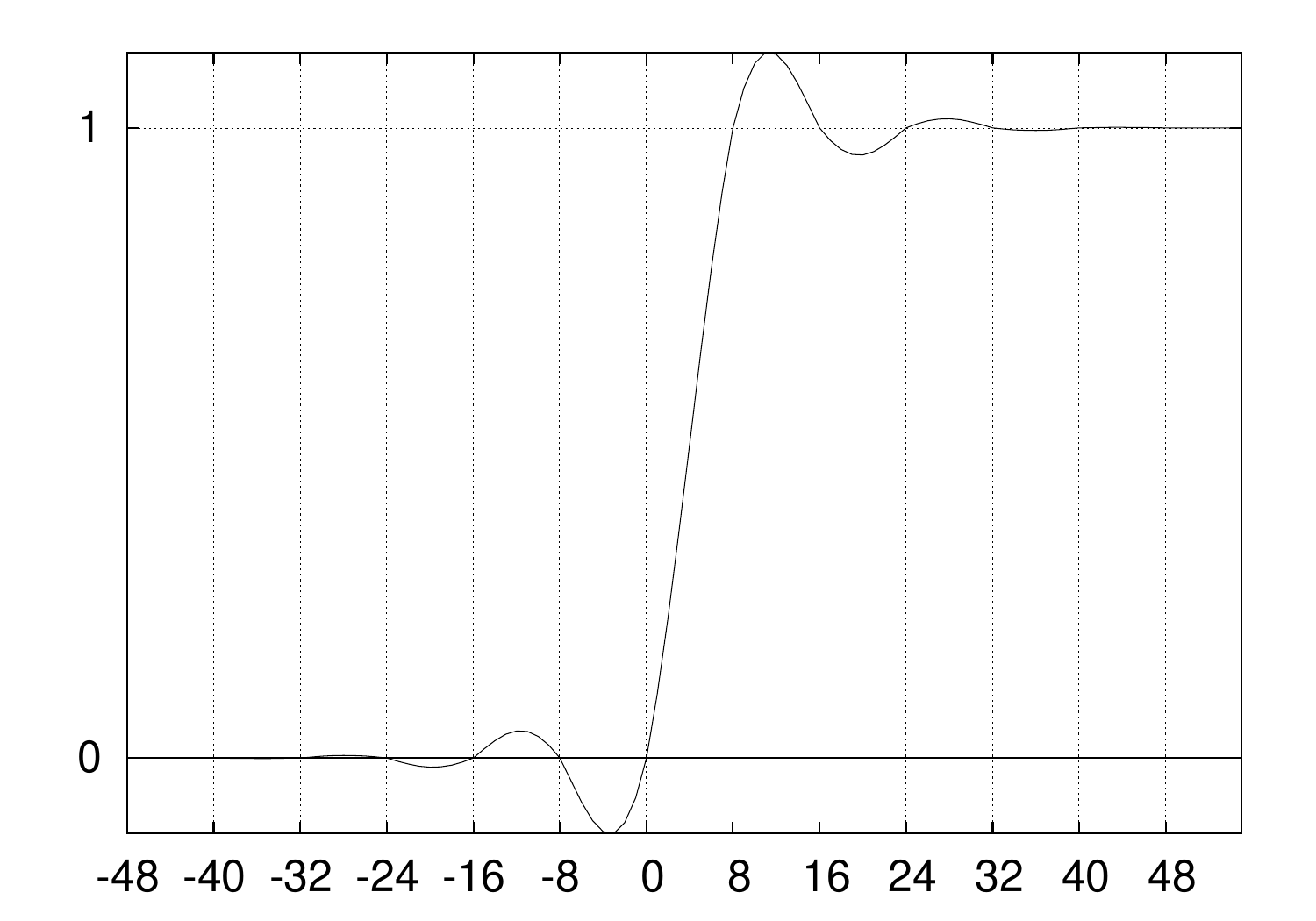}
 \end{center}
 \caption{\label{fig:wiggle}Given a step function $f$ going from $0$ to $1$,
 we show that the Lagrange interpolation $h$ is quite close to $f$ as we
 move away from the discontinuity. In this figure, $N=16$ and $b=8$.}
 \end{figure}

Recall subsection~\ref{section:piecewise}, in which we showed that
 $\sum f(kb)B_k$ was given by $\sum h(i)a_i$ where $h$ is a Lagrange
 interpolation of the range query function $f$. Only when
the function $f$ goes from a polynomial to 0 is there a
difference between the target range function $f$ and the range function $h$
estimated by bin-wise Lagrange interpolation.
However, the error made
by Lagrange interpolation also diminishes as we move away from the bin
containing the edge of the range. In many cases, it might be sufficient
to take into account only 1 or 3 bins near the edge (at each endpoint
of the range). For example, consider $N=4$, $b=1024$
with $f(x)= 1$ for $x> x_0$ and 0 otherwise.
A numerical evaluation shows that
using only one bin at each endpoint instead of the required 3 will
take care of 86\% of the error when range edges are in the middle
of a bin. The result is more significant for larger $N$, for example,
for $N=16$, we need 15 bins at each endpoint for a complete
evaluation; however, if we use only 5 centered, we take care of 97\%
of the error. (See Figure~\ref{fig:wiggle}.)
In short, \textsc{Ola} can provide wavelet-like
progressive evaluation of the queries simply by querying fewer bins in
the external array.

We can analyze more mathematically the relationship between the number
of bins used at each end of the range and the error.
First note that bin-wise Lagrange
interpolation is linear: if we interpolate the sequence
$\{0,0,0,1,0\}$ and the sequence $\{0,0,0,0,2\}$, then the sum of the
two interpolants is just the interpolation of the sequence
$\{0,0,0,1,2\}$.  Hence, it is sufficient to consider only one
non-zero sample value at any given time.
We proceed to show that the
contribution of a sample value $f(kb)$ to bin-wise Lagrange
interpolation decays quickly past one bin.  Let $N=M'+M$ be fixed,
and consider the interpolation of the sequence $x_0=1$, $x_{bi}=0$ for
all $i\neq 0$. We can then consider the interpolant $h$ in the
$k^{\mathrm{th}}$ bin defined by the interval $[bk,bk+b)$.
The following polynomial (refer to Eq.~(\ref{eq:lagrangeformula}))
describes $h$:
\begin{align*}
\frac{\prod_{i=M-1,i\neq k}^{M'} (x-kb+bi) }
{\prod_{i=M-1,i\neq k}^{M'} (-kb+bi) }
\end{align*}
where $x\in [bk,bk+b)$. The formula is only valid for $-M'\leq k\leq M-1$;
elsewhere $h$ is identically zero.
Clearly the denominator will increase sharply in absolute value as $k$ grows.
We show that the numerator is non increasing in $k$. Setting $y=x-kb\in [0,b]$,
we have $\prod_{i=M-1,i\neq k}^{M'} (x-kb+bi)=
\prod_{i=M-1, i\neq k}^{M'}(y+ib)$.
However, $\prod_{i=M-1, i\neq k}^{M'}(y+ib)=\frac{\prod_{i=M-1}^{M'}(y+ib)}{(y+kb)}$
and because $y\in [0,b]$, $y+kb \in [kb,kb+b]$ and so, the numerator goes down in amplitude as
$1/k$.
On the other hand, the denominator in absolute value,
$(b(-k+M-1))\cdots(b)(b)\cdots (b(k+M'))= b^{M+M'}(M-1-k)!(k+M')!$.
Setting $\lambda=k+M'$, we have
$b^{M+M'}(M-1-k)!(k+M')! =b^{M+M'}(M+M'-1-\lambda)!\lambda! = b^{M+M'}(N-1-\lambda)!\lambda!$
which has a rate of increase of starting at  $(N+2)/(N-2)$ and rising
with $k$.  Hence, the amplitude of the polynomial decreases faster than exponentially as
$k$ increases.

It is difficult to compare the progressive approximation we get using this
approach with related wavelet-based ones. Wavelet-based algorithms do not
use the original array as a data source when answering queries and thus, they have
much larger storage requirements. %We can say that f
For large-scale applications, approximate
queries are required for wavelet-based algorithms because storing all the coefficients
is unthinkable whereas \textsc{Ola} has progressive approximate queries as an added
option.

%%%%%%%%%%%%%%%%%%%%%%%%%%%%%%%%%%%%%%%%%%%%%%%%%%%%%%%%%%%%%%%%%%%%%%%%%%%%%%%%%%%%%%%%%%%%%%%%%%
\section{Conclusion and Future Work}
\label{sec:conclusion}
%%%%%%%%%%%%%%%%%%%%%%%%%%%%%%%%%%%%%%%%%%%%%%%%%%%%%%%%%%%%%%%%%%%%%%%%%%%%%%%%%%%%%%%%%%%%%%%%%%

\structure{Just a typical conclusion. }

This paper has considered bin-buffering algorithms and showed that using
a hierarchical approach, highly scalable %general
algebraic queries were possible
even with a small buffer. Using overlapped bins, we have shown that
we could buffer several local moments simultaneously and use
much less storage than  wavelet-based approaches while still supporting
progressive queries and very scalable queries and updates.

In short, we showed that $N$ local moments could be buffered using
only a single real-valued buffer: using bins of a fixed size irrespective
of $N$. %It is possible that
%owen: you have just made this stronger; do you actually know some?
Other
types of range queries could also be grouped and buffered
efficiently together~\cite{extendedwavelets}.
By a direct product~\cite{cascon}, Hierarchical Bin Buffering
and therefore the \textsc{Ola} approach
can be generalized to the multidimensional case.

Some implementation issues were not addressed. % in this paper.
For example,
many forms of buffering using finite-accuracy floating-point numbers
are susceptible to significant numerical errors.

Finally,
the source code used for the production of this paper is freely available~\cite{BinBufferingOnGoogle}.

\bibliographystyle{acmtrans}
\bibliography{1dbuffers}

\begin{thebibliography}{}

\bibitem[\protect\citeauthoryear{Alon, Matias, and Szegedy}{Alon
  et~al\mbox{.}}{1996}]{alon1996sca}
{\sc Alon, N.}, {\sc Matias, Y.}, {\sc and} {\sc Szegedy, M.} 1996.
\newblock The space complexity of approximating the frequency moments.
\newblock In {\em STOC'96}. ACM Press New York, NY, USA, 20--29.

\bibitem[\protect\citeauthoryear{Browne, Dongarra, Garner, Ho, and
  Mucci}{Browne et~al\mbox{.}}{2000}]{brow:papi-hpca}
{\sc Browne, S.}, {\sc Dongarra, J.}, {\sc Garner, N.}, {\sc Ho, G.}, {\sc and}
  {\sc Mucci, P.} 2000.
\newblock A portable programming interface for performance evaluation on modern
  processors.
\newblock {\em International Journal of High Performance Computing
  Applications\/}~{\em 14,\/}~3, 189--204.

\bibitem[\protect\citeauthoryear{Chakrabarti, Garofalakis, Rastogi, and
  Shim}{Chakrabarti et~al\mbox{.}}{2001}]{Approximatewavelets}
{\sc Chakrabarti, K.}, {\sc Garofalakis, M.}, {\sc Rastogi, R.}, {\sc and} {\sc
  Shim, K.} 2001.
\newblock Approximate query processing using wavelets.
\newblock {\em The VLDB Journal\/}~{\em 10,\/}~2-3, 199--223.

\bibitem[\protect\citeauthoryear{Cleveland and Loader}{Cleveland and
  Loader}{1995}]{cleveland-smoothing}
{\sc Cleveland, W.} {\sc and} {\sc Loader, C.} 1995.
\newblock Smoothing by local regression: Principles and methods.
\newblock Tech. rep., AT\&T Bell Laboratories.

\bibitem[\protect\citeauthoryear{Codd, Codd, and Salley}{Codd
  et~al\mbox{.}}{1993}]{olap}
{\sc Codd, E.~F.}, {\sc Codd, S.}, {\sc and} {\sc Salley, C.} 1993.
\newblock Providing {OLAP} ({On-line Analytical Processing}) to user-analysts:
  An {IT} mandate.
\newblock Tech. rep., E. F. Codd \& Associates.

\bibitem[\protect\citeauthoryear{Deligiannakis and Roussopoulos}{Deligiannakis
  and Roussopoulos}{2003}]{extendedwavelets}
{\sc Deligiannakis, A.} {\sc and} {\sc Roussopoulos, N.} 2003.
\newblock Extended wavelets for multiple measures.
\newblock In {\em SIGMOD}. ACM Press, 229--240.

\bibitem[\protect\citeauthoryear{Geffner, Agrawal, Abbadi, and Smith}{Geffner
  et~al\mbox{.}}{1999}]{geffner}
{\sc Geffner, S.}, {\sc Agrawal, D.}, {\sc Abbadi, A.~E.}, {\sc and} {\sc
  Smith, T.~R.} 1999.
\newblock Relative prefix sums: An efficient approach for querying dynamic
  {OLAP} data cubes.
\newblock In {\em ICDE'99}. 328--335.

\bibitem[\protect\citeauthoryear{Gray, Bosworth, Layman, and Pirahesh}{Gray
  et~al\mbox{.}}{1996}]{gray:datacube}
{\sc Gray, J.}, {\sc Bosworth, A.}, {\sc Layman, A.}, {\sc and} {\sc Pirahesh,
  H.} 1996.
\newblock Data cube: A relational aggregation operator generalizing group-by,
  cross-tabs and subtotals.
\newblock In {\em Proc, 1996 ICDE}. 131--139.

\bibitem[\protect\citeauthoryear{Ho, Agrawal, Megiddo, and Srikant}{Ho
  et~al\mbox{.}}{1996}]{ho97range}
{\sc Ho, C.-T.}, {\sc Agrawal, R.}, {\sc Megiddo, N.}, {\sc and} {\sc Srikant,
  R.} 1996.
\newblock Range queries in {OLAP} data cubes.
\newblock In {\em ACM SIGMOD}. 73--88.

\bibitem[\protect\citeauthoryear{IEC}{IEC}{1999}]{iec:new-nonSI-units}
{\sc IEC}. 1999.
\newblock Letter symbols to be used in electrical technology --- part 2:
  Telecommunications and electronics.
\newblock Tech. Rep. IEC 60027-2 Second Edition, International Electrotechnical
  Commission.

\bibitem[\protect\citeauthoryear{Jahangiri, Sacharidis, and Shahabi}{Jahangiri
  et~al\mbox{.}}{2005}]{1066189}
{\sc Jahangiri, M.}, {\sc Sacharidis, D.}, {\sc and} {\sc Shahabi, C.} 2005.
\newblock {SHIFT-SPLIT}: {I/O} efficient maintenance of wavelet-transformed
  multidimensional data.
\newblock In {\em SIGMOD '05}. 275--286.

\bibitem[\protect\citeauthoryear{Lemire}{Lemire}{2002}]{cascon}
{\sc Lemire, D.} 2002.
\newblock Wavelet-based relative prefix sum methods for range sum queries in
  data cubes.
\newblock In {\em CASCON'02}.

\bibitem[\protect\citeauthoryear{Lemire}{Lemire}{2007}]{lemire2007}
{\sc Lemire, D.} 2007.
\newblock A better alternative to piecewise linear time series segmentation.
\newblock In {\em SDM'07}.

\bibitem[\protect\citeauthoryear{Lemire and Kaser}{Lemire and
  Kaser}{2007}]{BinBufferingOnGoogle}
{\sc Lemire, D.} {\sc and} {\sc Kaser, O.} 2007.
\newblock Hierarchical bin buffering library in {C++}.
\newblock \url{http://code.google.com/p/hierarchicalbinbuffering/}, last
  checked on 15/7/2007.

\bibitem[\protect\citeauthoryear{Li and Shen}{Li and Shen}{1992}]{liimageseg}
{\sc Li, B.-C.} {\sc and} {\sc Shen, J.} 1992.
\newblock Fast calculation of local moments and application to range image
  segmentation.
\newblock In {\em Int. Conf. Pattern Recognition}. 298--301.

\bibitem[\protect\citeauthoryear{Moerkotte}{Moerkotte}{1998}]{moerkotte98}
{\sc Moerkotte, G.} 1998.
\newblock Small materialized aggregates: A light weight index structure for
  data warehousing.
\newblock In {\em VLDB'98}. 476--487.

\bibitem[\protect\citeauthoryear{Patterson}{Patterson}{2003}]{grayqueue}
{\sc Patterson, D.} 2003.
\newblock A conversation with {Jim Gray}.
\newblock {\em ACM Queue\/}~{\em 1,\/}~4 (June), 6--7.

\bibitem[\protect\citeauthoryear{Poon}{Poon}{2003}]{poonortho2003}
{\sc Poon, C.} 2003.
\newblock Dynamic orthogonal range queries in {OLAP}.
\newblock {\em Theoretical Computer Science\/}~{\em 296,\/}~3, 487--510.

\bibitem[\protect\citeauthoryear{Rao}{Rao}{2002}]{rao2002}
{\sc Rao, S.~S.} 2002.
\newblock {\em Applied Numerical Methods for Engineers and Scientists}.
\newblock Prentice Hall.

\bibitem[\protect\citeauthoryear{Schmidt and Shahabi}{Schmidt and
  Shahabi}{2002}]{propolyne}
{\sc Schmidt, R.~R.} {\sc and} {\sc Shahabi, C.} 2002.
\newblock Propolyne: A fast wavelet-based algorithm for progressive evaluation
  of polynomial range-sum queries.
\newblock In {\em Conference on Extending Database Technology}. 664--681.

\bibitem[\protect\citeauthoryear{Scott and Sagae}{Scott and
  Sagae}{1997}]{binscott1997}
{\sc Scott, D.} {\sc and} {\sc Sagae, M.} 1997.
\newblock Adaptive density estimation with massive data sets.
\newblock In {\em ASA, Statistical Computing Section}. 104--108.

\bibitem[\protect\citeauthoryear{Silva, Chiang, El-Sana, and Lindstrom}{Silva
  et~al\mbox{.}}{2002}]{silva02outcore}
{\sc Silva, C.}, {\sc Chiang, Y.}, {\sc El-Sana, J.}, {\sc and} {\sc Lindstrom,
  P.} 2002.
\newblock Out-of-core algorithms for scientific visualization and computer
  graphics.
\newblock In {\em Visualization'02 Course Notes}.

\bibitem[\protect\citeauthoryear{Vitter}{Vitter}{2002}]{vitter2002externalalgo%
s}
{\sc Vitter, J.~S.} 2002.
\newblock {\em Handbook of massive data sets}.
\newblock Kluwer Academic Publishers, Chapter External memory algorithms,
  359--416.

\bibitem[\protect\citeauthoryear{Vitter, Wang, and Iyer}{Vitter
  et~al\mbox{.}}{1998}]{Vitterdatacubeapprox}
{\sc Vitter, J.~S.}, {\sc Wang, M.}, {\sc and} {\sc Iyer, B.} 1998.
\newblock Data cube approximation and histograms via wavelets.
\newblock In {\em CIKM}. ACM Press, 96--104.

\bibitem[\protect\citeauthoryear{Zhou and Kornerup}{Zhou and
  Kornerup}{1995}]{zhou95computing}
{\sc Zhou, F.} {\sc and} {\sc Kornerup, P.} 1995.
\newblock Computing moments by prefix sums.
\newblock Tech. Rep. PP-1995-31, University of South Denmark.

\end{thebibliography}

\end{document}